\documentclass[12pt]{iopart}
\usepackage[utf8]{inputenc}
\usepackage{calc}
\usepackage{graphicx}
\usepackage{array} 
\usepackage{color}

\newcommand{\text}[1]{{\rm #1}}
\definecolor{Red}{rgb}{0.9,0.1,0.1}
\definecolor{Blue}{rgb}{0.1,0.0,0.9}
\definecolor{Darkblue}{rgb}{0.22,0.33,0.64}

\begin{document}
\title{Effects of microtubule mechanics on  hydrolysis and catastrophes}

\author{N M\"uller and J Kierfeld}

\address{Department of Physics, TU Dortmund University, 44221 Dortmund,
  Germany}
\ead{nina2.mueller@tu-dortmund.de, jan.kierfeld@tu-dortmund.de}

\begin{abstract}
We introduce a 
model for microtubule  mechanics 
containing lateral bonds between dimers 
in neighboring protofilaments,
bending rigidity of dimers, and repulsive interactions 
between protofilaments modeling steric constraints to
investigate the influence of mechanical forces on hydrolysis 
and catastrophes. 
We use the allosteric dimer model, where
 tubulin dimers are characterized by an equilibrium bending angle,
which changes from $0^\circ$ to 
$22^\circ$ by hydrolysis of a dimer. 
This also  affects the lateral interaction and bending 
energies and, thus, the mechanical equilibrium state of the microtubule. 
As hydrolysis gives rise  to conformational changes in  dimers, 
mechanical forces also influence the hydrolysis rates by mechanical 
energy changes modulating the hydrolysis rate. 
The interaction via the  microtubule mechanics
then gives rise to  correlation effects in the hydrolysis 
dynamics, which have not been taken into account before. 
Assuming a dominant influence of  mechanical energies on hydrolysis 
rates, we investigate the most probable 
 hydrolysis pathways both for vectorial and 
random hydrolysis. 
Investigating the  stability with respect to lateral bond rupture,
we identify   initiation  configurations
for catastrophes along the hydrolysis pathways and 
values for a lateral bond rupture force. 
If we allow for rupturing of lateral bonds between dimers 
in neighboring  protofilaments above this threshold force,
our model exhibits avalanche-like catastrophe events. 
\end{abstract}


\submitto{\PB}


%%%%%%%%%%%%%%%%%%%%%%%%%
\section{Introduction}

Microtubule (MT) dynamics is essential for many cellular processes, 
such as 
 cell division \cite{Mitchison2001}, 
  intracellular positioning processes \cite{D05}, 
e.g.\  positioning of 
 the cell nucleus \cite{daga2006} or chromosomes during
mitosis, establishing cell polarity \cite{SD07}, or regulation of
cell shapes \cite{Picone2010,Dehmelt2003}. 
An important feature of MT dynamics is their dynamic 
instability, which is the  stochastic switching between phases 
of growth and rapid shrinkage \cite{Mitch1984}.  
Polymerization phases terminate in catastrophes, where the 
MT switches to a state of rapid depolymerization. 
Depolymerization phases are terminated by rescue events, 
in which the MT switches back into a growing state.  
The dynamic instability is closely linked to the hydrolysis 
of tubulin dimers:   MT catastrophe events 
are associated with the loss of the stabilizing GTP-cap 
through hydrolysis to GDP-tubulin within the MT. 

The depolymerization rate of GDP-tubulin dimers is significantly higher
than the depolymerization rate of GTP-dimers.
However, catastrophes are more than 
phases of rapid depolymerization of the GDP-cap of a MT.
This is strongly suggested by conformational changes 
and mechanical forces occurring during catastrophes:
(i) in a MT catastrophe the protofilaments of MT 
fall apart and curl into  characteristic ``ram's horn'' conformations
 \cite{Kirschner1974};
(ii) depolymerizing MTs can exert forces, which are 
 important in mitosis if chromosomes are separated \cite{Mitchison2001}.

The curling into ram's horns is caused by a spontaneous curvature 
of GDP-tubulin dimers. 
This means hydrolysis gives rise to changes in the tubulin dimer 
conformation or the MT structure.
Two different models have been discussed in the literature to 
describe the influence of hydrolysis on the mechanics of the MT, 
the allosteric model \cite{Wang2005}  and the lattice model
\cite{Buey2006,Wu2012}. 
In the allosteric model,
inter-dimer or intra-dimer bending angles change during hydrolysis 
\cite{Wang2005,nogales2006}.
Hydrolysis of a tubulin dimer changes one or both of these angles 
\cite{Wang2005}. If it is assumed that hydrolysis mainly changes 
the intra-dimer angle, this
angle increases from  $0^\circ$ for straight GTP-dimers
to $22^\circ$ for curved GDP-dimers   \cite{Mueller1998}.
In the lattice model,
both states of the tubulin dimers are slightly bent and hydrolysis 
weakens the lateral interaction strength between dimers in neighboring
protofilaments \cite{Wang2005,Buey2006,Wu2012}. 
At present, 
the experimental evidence is not sufficient to rule out any of the two
models.
Also combinations of both models are possible, where hydrolysis 
affects both the intra-dimer angle (allosteric) and the  
interaction strength between laterally neighboring dimers (lattice).

The influence of tubulin dimer hydrolysis onto the mechanics of 
the MT lattice suggests
that, vice versa, mechanical forces and torques 
acting on  tubulin dimers or the MT structure 
also affect hydrolysis rates. 
This effect has not been considered in the literature before,
and
we will use the allosteric model with intra-dimer bending 
to study this coupling of hydrolysis and MT mechanics. 
Similar investigations should be done in the future for 
the lattice model.

Moreover, recent experiments also suggest that catastrophes are initiated
in a multi-step process, which involves not a single 
 rate-limiting event but 
a chain of at least two events, which are probably related to 
hydrolysis  events  \cite{Gardner2011a}.
  Mechanical forces might be one possible way 
to orchestrate such a chain of hydrolysis events. 
Our results on the influence of MT mechanics on the hydrolysis 
pathway will give hints on the mechanism of catastrophe initiation.

There are numerous models for the growth dynamics of MTs,
which either ignore catastrophe events and focus on the 
growing phase of MT dynamics 
\cite{Doorn2000,Kolomeisky2001,Stukalin2004,Ranjith2009,Krawczyk2011}, 
or which include catastrophes as explicit stochastic switching events on a
macroscopic level following \cite{Dogterom1993}. These approaches
include 
details of the hydrolysis mechanism into the 
model for the catastrophe rate 
\cite{Flyvbjerg1994,Flyvbjerg1996,Zelinski2012,Zelinski2013}. 
One  focus of  these  approaches was the explanation of the behavior 
of MT growth dynamics under force.

In this paper,  
we want to concentrate on  the coupling of 
hydrolysis events  to MT mechanics.
We will ignore the polymerization dynamics for simplicity and consider 
hydrolysis in MTs of fixed length without any external forces. 
In the end, we  try to develop a microscopic model for 
the initiation of catastrophes 
based on hydrolysis events coupled to the mechanics of the MT lattice
and additional rupture events within the MT lattice. 
This means that we will not include catastrophes 
as explicit stochastic events but obtain catastrophes as emerging 
events from hydrolysis coupled to mechanics within the MT. 
Our description will  be microscopic on the level of 
tubulin dimers (i.e., not on the level of all-atom 
simulations as in \cite{Sept2010,Grafmuller2011,Grafmuller2013}).

There exist already various models 
describing the stochastic dynamics of  hydrolysis 
and the mechanics of MTs 
 on the dimer level. They can be classified into different 
categories:

(i) There are purely ``chemical'' models not taking into account 
the mechanics of the MT lattice but  only including 
chemical rate constants for addition and removal of tubulin dimers and 
hydrolysis rates
\cite{Flyvbjerg1994,Flyvbjerg1996,Bayley1990,VanBuren2002,Brun2009a,
Bowne2013,Jemseena2013,Padinhateeri2012,Li2013}. 
 In these chemical models, 
phases of accelerated depolymerization can be observed if the 
cap mainly consists of GDP-tubulin. These phases of 
fast depolymerization are identified with catastrophe events. 
Different hydrolysis mechanisms such as 
random \cite{VanBuren2002,Padinhateeri2012}, 
vectorial \cite{Bayley1990,Ranjith2009}, 
or mixed cooperative mechanisms
\cite{Flyvbjerg1994,Flyvbjerg1996,Li2009,Li2013} are possible and 
have been discussed, but this issue 
is not settled for microtubules \cite{Bowne2013}. 
The model proposed by Flyvbjerg  \cite{Flyvbjerg1994,Flyvbjerg1996}
including  both random and vectorial mechanisms in a cooperative
mechanism  has been successfully applied to fit catastrophe rates from 
the resulting first-passage statistics to zero GTP-cap length. 
The results suggest a strongly cooperative mechanism with mostly
vectorial hydrolysis and a fairly small random component. 
However, if the results of \cite{Li2010} for the resulting 
GTP-cap length are used, the strong cooperativity leads to rather 
large GTP-caps as compared to experimental findings of up to 
3 layers of GTP-tubulin \cite{Drechsel1994,Desai1997,Schek2007}.
In Ref.\ \cite{Li2013}, on the other hand, 
a weakly cooperative mechanism with mostly random hydrolysis is found
 to describe experimental data on catastrophe rates best. 
In \cite{Schek2007}, 
a random hydrolysis mechanism has been 
successfully used to model typical MT length fluctuations on the 
nanometer scale. 
In \cite{Bowne2013}, a coupled-random hydrolysis mechanism 
has been proposed, 
where the plus end GTP-dimer  cannot hydrolyze but only dissociate.  
One advantage of  many  chemical models is that 
exact or, at least, approximative analytical results can be found
\cite{Flyvbjerg1994,Flyvbjerg1996,Li2009,Padinhateeri2012,Li2013}, 
for example, for catastrophe frequencies, which can be
easily  compared  to experimental data.

(ii) Chemical models also differ with respect 
to protofilament substructure. In some 
models  the 13 protofilament substructure  is ignored
\cite{Flyvbjerg1994,Flyvbjerg1996,Ranjith2009,Padinhateeri2012,Li2013}, 
more detailed models contain all 13 protofilaments
explicitly \cite{Bayley1990,VanBuren2002,Brun2009a,Bowne2013,Jemseena2013}.
Some of these more detailed ``chemical'' models also include
a possible  influence
of the neighboring protofilament state on hydrolysis 
\cite{VanBuren2002,Brun2009a}.
However, all of these models 
 still deal with strictly rigid MTs.

(iii) Furthermore, there are 
purely ``mechanical''  models on the level of single dimers, which 
 include   the possible 
bending of protofilaments and binding interaction between 
neighboring protofilaments
\cite{Sept2010,Grafmuller2011,Grafmuller2013,Molodtsov2005,Mohrbach2010}, 
but do not model their influence 
onto chemical hydrolysis or polymerization rates.

(iv) Only the model proposed in \cite{VanBuren2005} 
couples chemical and mechanical aspects of the MT lattice and 
considers the influence of mechanics on polymerization 
and depolymerization rates.

In all of these models, 
the interplay of mechanical forces and hydrolysis remains
unaddressed so far.  
In this paper, we will add  this aspect and
 focus on the influence of mechanical forces onto hydrolysis
rates and its consequences for  hydrolysis pathways and the initiation of 
catastrophe events.

%%%%%%%%%%%%%%%%
\section{Methods}

%%%%%%%%%%%%%%%%%%%%%%%%%%%%%%%%
\subsection{Mechanical microtubule model}

Each protofilament consists of $\alpha\beta$-tubulin heterodimers 
of length $d=8{\rm nm}$.
MTs consist of 13 protofilaments forming a hollow tube
of (outer) radius $R_o=12.5{\rm nm}$.
GTP can bind to the  $\beta$-tubulin and is hydrolyzed in the 
polymerized state. 

In the following, we will employ the so-called allosteric model 
and assume that GTP-tubulin dimers are straight and 
assemble into straight 
protofilaments, whereas GDP-tubulin tends to form curved protofilaments
with curvature radius $21{\rm nm}$ in the typical ram's horn 
configurations \cite{Mueller1998}.
Protofilament curvature can be caused both by inter- and intra-dimer 
bending \cite{Wang2005,nogales2006}. 
If we assume that there is only intra-dimer bending during 
hydrolysis, a GDP-dimer acquires a bent configuration 
with an equilibrium angle of $22^\circ$. 
The tube formed by the 13 protofilaments is stabilized by 
lateral bonds between tubulin dimers in neighboring protofilaments. 
These bonds  are assumed to be identical for GTP- and GDP-dimer
within the allosteric model.

GTP-tubulin assembles into protofilaments (i.e., has a high 
polymerization rate and a low depolymerization rate), whereas 
GDP-tubulin tends to disassemble (i.e., has a depolymerization 
rate much higher  than the polymerization rate) \cite{Desai1997}. 
Within the MT protofilaments, GTP-tubulin is hydrolyzed 
into GDP-tubulin after a certain waiting time depending on the 
exact hydrolysis mechanism and hydrolysis rates. 
This gives rise to MTs consisting of an unhydrolyzed  GTP-cap 
at the growing end and hydrolyzed GDP-tubulin behind this cap. 
Comparison of recent 
experimental measurements of MT length fluctuations
on the nanometer scale 
and a random hydrolysis model 
suggests a cap size  of $\sim 40$ GTP-dimers
corresponding  to 3 layers of GDP-dimers 
\cite{Schek2007,VanBuren2005}.
The structure of the cap will depend on the hydrolysis mechanism.
For  {\it random} hydrolysis, where 
dimers hydrolyze completely independent from each other 
\cite{Mitch1984,VanBuren2005},
the cap boundary is not sharp (in the sense 
that there are many GTP-/GDP-boundaries in the cap region) 
with  an exponentially decaying  GTP-tubulin concentration.
For a   {\it vectorial} mechanism, 
where hydrolysis propagates as a ``wave'' because 
hydrolysis can only happen if the neighboring dimer is 
already in its hydrolyzed GDP-state \cite{Carlier1987},
the cap boundary is sharp by definition. 
For {\it cooperative} models incorporating both mechanisms
\cite{Flyvbjerg1994,Flyvbjerg1996,Li2009}, we get a 
cap structure consisting of many GTP-islands with a 
characteristic dependence of the island size distribution 
 and cap size on the 
cooperativity parameter \cite{Li2009,Li2010}.

Because GTP-tubulin dimers are straight, they stabilize the MT
as they do not stress lateral bonds between protofilaments.
Intrinsically curved GDP-dimers, on the other hand, 
 stress the lateral bonds and destabilize the MT lattice.  
In the following,  we consider a MT with a 
fixed length, a stabilizing 
GTP-cap of fixed size and with a sharp cap 
boundary for simplicity (typically 3 layers of GTP-dimers). We
formulate a mechanical model, which 
(i)  incorporates dimer bending and 
lateral bonds between neighboring protofilaments, 
(ii) gives mechanically stable tubular structures
in agreement with experimentally observe MT 
structures, and 
(iii) is as simple as possible.

\begin{figure}
 \includegraphics[width=\textwidth]{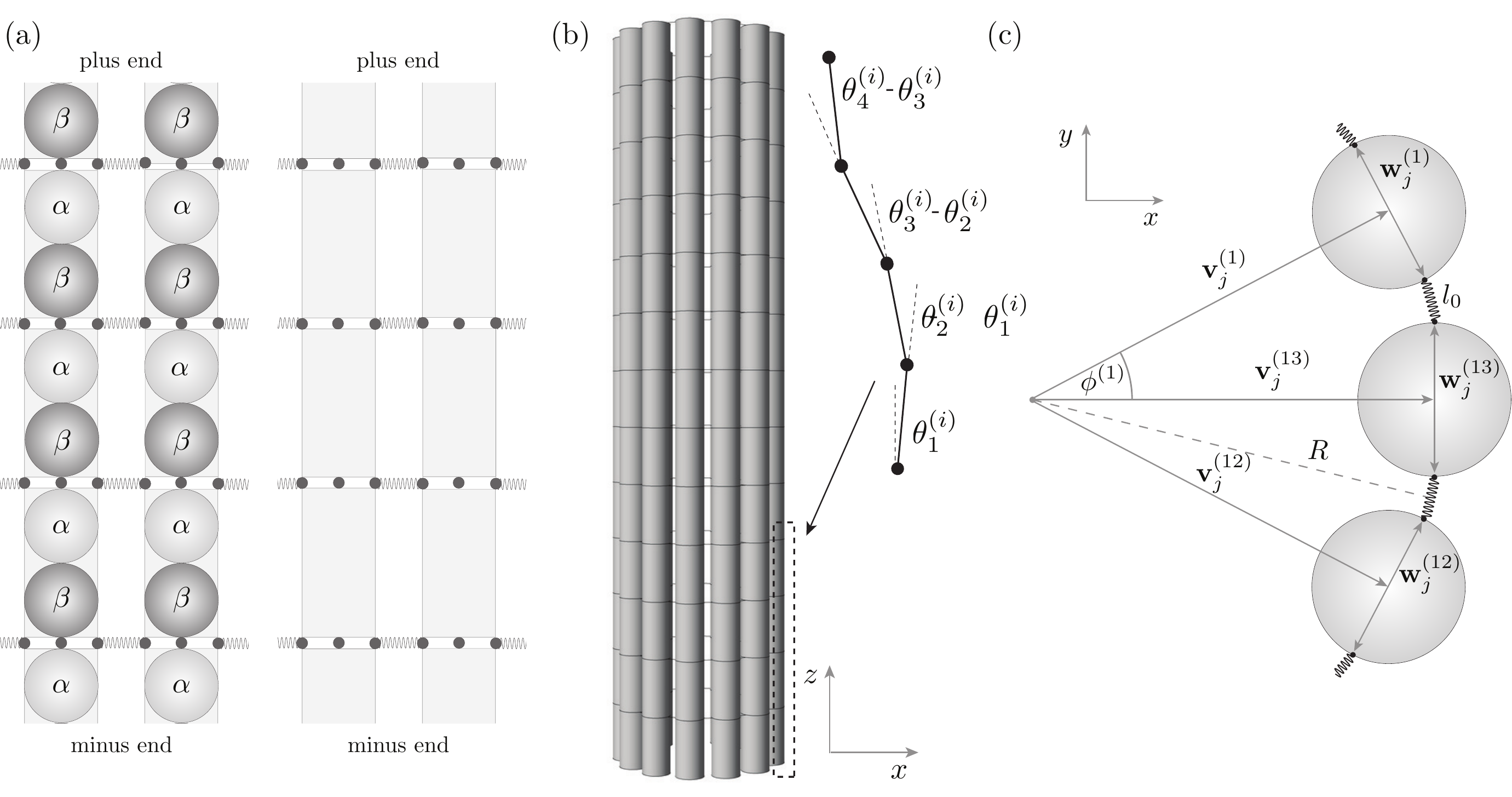}
 \caption{
   Schematic representation of the MT model.
  (a) Only the  $\alpha\beta-$ junction of tubulin dimers can bend. 
    Straight
    $\beta\alpha$-segments are grouped to stiff rods of length $d$ and 
   diameter $d/2$ for the mathematical
   description. Each rod has two interaction points for 
   lateral bond springs 
   at the surface of the rods
   at mid-height of the dimers, which means at the top end of the rods. 
  (b)
   A specific configuration is characterized by the position angles
   $\theta_j^{(i)}$. 
     Hydrolysis causes a shift of the preferred bending
   angle $\theta_{0,j}^{(i)}$ 
  from $0^\circ$ to $22^\circ$ at the $\alpha\beta-$ junction. 
   (c) Lateral bonds are modeled by spring interactions. The
   interaction points lie on the surface of the tubulin dimers.}
 \label{fig:MT_def}
\end{figure}

 MTs are hollow cylinders consisting of 13 protofilaments with 
 outer radius $R_o=12.5{\rm nm}$ and inner radius $R_i=8.5{\rm nm}$,
where we assume each tubulin 
dimer to consist of two spherical tubulin monomers of size $d/2 = 4{\rm nm}$. 
 Each protofilament is modeled 
by a chain of $M+1$   $\alpha\beta$-tubulin dimers, 
which are longitudinally bound by intra-dimer and inter-dimer bonds. 
The total length of the MT is $L=(M+1) d$ with the dimer length $d$. 
We only allow for intra-dimer bending 
 at the $\alpha\beta$-junction within each dimer, see figure \ref{fig:MT_def}a.
The MT  conformation is then described by
bending angles of the $\alpha\beta$-dimer 
in  the $i^{\rm th}$ protofilament ($i=1,...,13$)
 in the $j^{\rm th}$ layer, where $j=M$ corresponds to the 
GTP-capped MT end and $j=1$ to the GDP-end, see figure \ref{fig:MT_def}b.
We only allow  for intra-dimer 
bending  and regard the inter-dimer bonds as straight and 
fixed. 

For the mathematical description, we can then group 
the $\beta$-tubulin in layer $j-1$ with the $\alpha$-tubulin 
in layer $j$ into a straight ``rod'' $j$ oriented with angle 
$\theta_j^{(i)}$ in protofilament $i$. In the following we will describe 
the MT configuration in terms of these rods, as shown 
in figure \ref{fig:MT_def}a.
We fix  $\theta_0^{(i)}=0$ and parametrize the MT configuration 
by $M$ angles $\theta_j^{(i)}$ ($j=1,...,M$) for each protofilament $i$.
Vice versa, the  bendable junction between 
rods $j-1$ and $j$ in protofilament $i$ 
corresponds to the bendable $\alpha\beta$-junction of the dimer $j$
with bending angle 
$\Delta \theta_j^{(i)} = \theta_{j}^{(i)}-\theta_{j-1}^{(i)}$. 
We note that
 the last $\beta$-tubulin monomer of the last tubulin dimer 
 at the plus end in layer $j=M$ is  not contained in any rod.
Therefore, the hydrolysis state of the last tubulin dimers 
 at the plus end of each protofilament will have no effect in our model,
as further discussed below.

\begin{table}
  \caption{\label{tab:geometry} 
   MT geometry parameters.}
\begin{indented}
\item[]
\begin{tabular}{@{}ll}
\br
 dimer size $d$ & 8nm \\
 monomer size $d/2$ & 4nm\\
 outer radius $R_o$ & 12.5nm\\
 inner radius $R_i$ & 8.5nm\\ 
  mean radius $R$  & 10.5nm\\
 spring length $l_0$ & 1.14nm\\
 protofilament distance $r$ & 5.1nm\\
 protofilament shift $z_0$ & 0.92nm\\
\br
\end{tabular}
\end{indented}
\end{table}

 The minus end of 
every protofilament is fixed in the $xy$-plane. 
For simplicity, we do not take into
account the helical pitch. 
We choose the MT axis as $z$-axis and define
\begin{eqnarray}
 {\bf v}_0^{(i)}=\left( 
		    \begin{array}{@{}c@{}} 
		      R\cos\phi^{(i)}\\
		      R\sin\phi^{(i)}\\
		      0 
		    \end{array}\right)
\end{eqnarray}
as starting points of the first layer of rods. 
The polar angle of the $i^{\rm th}$ protofilament is
 $\phi^{(i)}=i\cdot 2\pi/13$. 
Thus, the ending position ${\bf v}_{j}^{(i)}$ of the $i^{\rm th}$ rod 
in the $j^{\rm th}$ layer is given by
\begin{eqnarray}
 {\bf v}_{j}^{(i)}={\bf v}_{j-1}^{(i)}+d\left( 
		    \begin{array}{@{}c@{}} 
                \cos\phi^{(i)}\sin\theta_j^{(i)}\\
		\sin\phi^{(i)}\sin\theta_j^{(i)}\\
		\cos\theta_j^{(i)}
                 \end{array}\right). \label{eq:vi}
\end{eqnarray}
We assume that dimers or 
rods can only be displaced in radial direction, i.e., 
all polar  angles  $\phi_j^{(i)} = \phi^{(i)}= i\cdot 2\pi/13$ are fixed  and
independent of the layer number $j$. 
The azimuth angles $\theta_j^{(i)}$ 
of rods describe the radial displacements of dimers and 
are the configurational variables that
are determined by mechanical energy minimization.

We consider MT configurations that contain $m_1$ complete layers of
hydrolyzed GDP-tubulin and $m_2+1$ complete layers of GTP-tubulin
($M+1=m_1+m_2+1$); we focus on  4 layer GTP-caps
with  $m_2=3$.
 In accordance to the allosteric model we assume 
GTP-dimers to be straight with an equilibrium angle  $\theta_0=0^\circ$
and GDP-tubulin bent with an equilibrium angle  $\theta_0=22^\circ$.
We consider deviations from these preferred equilibrium
angles to be governed by an elastic bending energy with a 
characteristic bending rigidity $\kappa$.
Therefore, we  define the longitudinal bending  energy stored
in the MT lattice as
\begin{eqnarray}
 E_{\rm long}=E_{\rm bend}= 
  \sum_{j=1}^{M}\sum_{i=1}^{13} \frac{\kappa}{2} 
   \left( \theta_j^{(i)}-\theta_{j-1}^{(i)}-\theta_{0,j}^{(i)}\right)^2, 
\label{eq:long_int}
\end{eqnarray}
with $\theta_{0}^{(i)}=0$ fixed and the bending rigidity $\kappa$ 
of individual  tubulin dimers.

To obtain a tubular structure, we have to introduce 
lateral bonding. A reasonable and simple assumption is that
 only dimers in neighboring protofilaments
interact laterally (reminiscent of some form of molecular bonds 
forming). Mechanical stability will 
require to introduce two contributions to these  lateral interactions:
one attractive binding contribution in form of a harmonic  bond with a 
finite rest length and  an additional hard-core-like repulsion,
which basically prevents configurations with 
interpenetrating dimers.

For the attractive binding contribution, 
we assume that neighboring tubulin dimers interact via bonds 
between specific interaction points on the surface of the dimers.
We use only one lateral bond per dimer, i.e., each 
dimer or each rod has two interaction points, which we locate
 on the surface  at the top
of each rod, see figure \ref{fig:MT_def}a.
We model tubulin dimers and rods 
with a  spherical cross section  of radius $d/4$ and 
assume that the  bonds between neighboring dimers have a 
rest length $l_0$, see figure \ref{fig:MT_def}c. We consider lateral
stretching from the rest length $l_0$ to be harmonic with a 
characteristic spring constant $k$.
The rest length $l_0$ can be determined by geometry; with 
the  mean  radius $R=(R_o+R_i)/2 = 10.5{\rm nm}$  and assuming 
spherical tubulin monomers of radius $d/4$, we obtain 
\begin{equation}
   l_0 = 2 (\sin \pi/13)[ R-(d/{4}) (\cot \pi/13)] \simeq 1.14 {\rm nm}.
\label{eq:l0}
\end{equation}

If the equilibrium angle $\theta_{0,j}^{(i)}$ of a tubulin dimer changes
from $0^\circ$ to $22^\circ$ by hydrolysis, this will strain the 
lateral bonds in the MT lattice.
Because the  interaction points  for lateral bonds are  located
 at the top of each rod (or at mid-height of each dimer, see figure 
\ref{fig:MT_def}a), 
 hydrolysis of the  last $\beta$-tubulin monomer of the last tubulin dimer 
 at the plus end ($j=M$) does {\em not} strain the last bond. 
Therefore,  the hydrolysis state of the  tubulin dimers 
 at the plus end of each protofilament at $j=M$ will have no effect on 
MT mechanics in our model.
If we consider MT-configurations with GTP-caps 
consisting of 
$m_2+1$ complete layers of GTP-tubulin, only 
 $m_2$ GTP-layers have an effect  on MT mechanics. We therefore
ignore the last   layer of $\beta$-tubulin monomers at the plus end 
in the following, in particular in 
the MT hydrolysis patterns  shown in 
figures \ref{fig:HP_random_05} -- \ref{fig:HP_vec_0005} below.
We point out that the last tubulin layer has a stabilizing effect 
also in our model as the stabilizing lateral bonds are contained 
in our model (at the top of the last rod, see figure \ref{fig:MT_def}a). 
However, this stabilizing effect is independent  
of the hydrolysis state of the last layer and only depends on the 
hydrolysis states of the layers below. 
This point of view can be further  justified because 
 GTP-dimers at the plus end 
 cannot hydrolyze as the last $\beta$-monomer 
has no  inter-dimer contact \cite{Nogales2000}.

In addition to the attractive lateral binding forces, 
we apply a strongly repulsive hard core interaction 
to avoid overlapping
of tubulin dimers, which we model by a $r^{-12}$-potential. 
Accordingly, we define the total lateral mechanical energy by
\begin{eqnarray}
 E_{\rm lat}&=&E_{\rm spring}+E_{\rm hc} \nonumber \\
	     &=&\sum\limits_{j=1}^{M}\sum\limits_{i=1}^{13} 
   \frac{k}{2}\left[\left|{\bf v}_j^{(i)}-{\bf w}^{(i)}
  -({\bf v}_j^{(i-1)}+{\bf w}^{(i-1)})\right|-l_0\right]^2  \nonumber \\
	     &&+\sum\limits_{j=1}^{M}\sum\limits_{i=1}^{13} 
   k'\cdot\left( \left|{\bf v}_j^{(i)}-{\bf v}_j^{(i-1)}\right|
    -{\frac{d}{2}} \right)^{-12},
\end{eqnarray}
where $\pm {\bf w}^{(i)}$ are the vectors from the 
center of the spherical dimer cross section to the two 
  interaction points on the dimer surface,
$k$ is the  spring constant for the binding force 
and  $k'$   the strength of the repulsion.

As the stable configuration of a physical system is characterized by minimum
free energy, we numerically calculate the minimum of the total 
energy functional 
\begin{equation}
  E =  E_{\rm long}+ E_{\rm lat} = 
    E_{\rm bend} + E_{\rm spring}+E_{\rm hc}
\end{equation}
with respect to the $13\cdot (m_1+m_2)$ variables $\theta_j^{(i)}$.

In the following, we often measure 
 energies in units of the dimer bending rigidity 
$\kappa$. This shows that the dimensionless energy $E/\kappa$ 
and, thus, our  MT model is characterized by three 
remaining control parameters:
the ratios $k/\kappa$ and $k'/\kappa$ 
control the MT mechanics;
the ratio  $\kappa/k_BT$ of dimer bending rigidity and 
thermal energy controls the 
 relevance of stochastic thermal  fluctuations
in comparison to mechanical forces.
All relevant model parameters for MT geometry are summarized in table 
  \ref{tab:geometry}.

%%%%%%%%%%%%%%%%%%%%%%%%%%%%%%%%%%%%%%%%%%%
\subsection{Model parameters for mechanically stable 
  tubular structures}

No direct experimental measurement of the  microscopic
mechanical parameters  $k$ and $\kappa$ is available so far. 
However, there are a number of 
experiments determining  the macroscopic elastic moduli 
of the MT lattice which could be related to the 
microscopic mechanical parameters. 
The lateral bond elasticity $k$ can be deduced from 
measurements of the macroscopic shear modulus \cite{Sept2010}. 
Moreover, molecular dynamics (MD)  simulations of short MT protofilaments 
 \cite{Sept2010,Wells2010,Grafmuller2011,Grafmuller2013} can eventually 
give more direct information on the molecular scale parameters
$\kappa$ and  $k$.
The requirement to obtain a stable equilibrium MT 
structure, which is tubular and agrees with the 
experimentally observed geometry 
will put further constraints on the three 
 parameters $\kappa$, $k$, and $k'$
of our model.

\begin{figure}
  \includegraphics[width=0.5\textwidth]{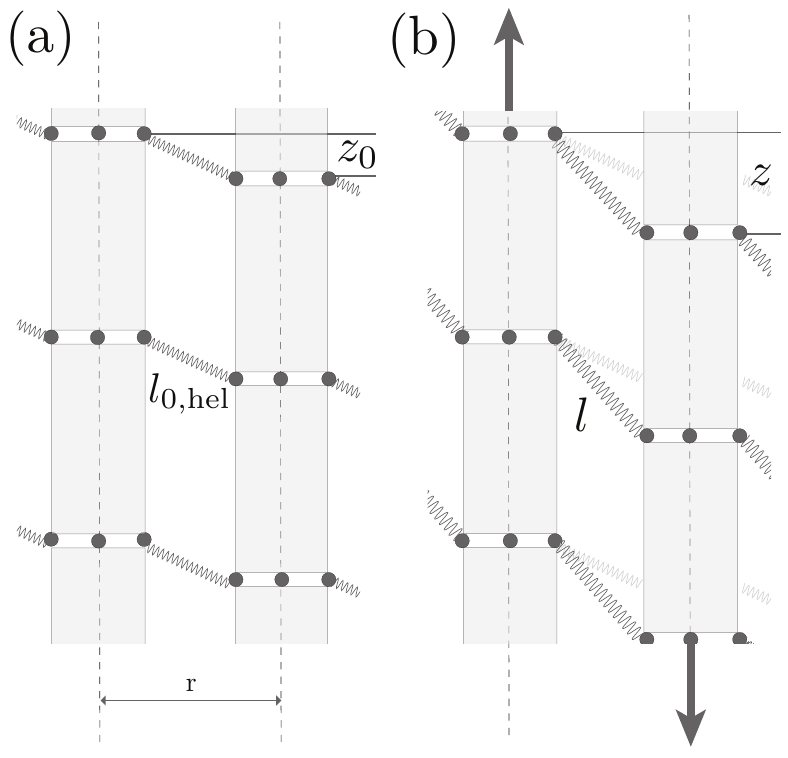}
  \caption{Deformation of lateral bond springs by
   shearing the MT lattice.}
\label{fig:shear}
\end{figure}

The lateral bond elasticity
$k$  can be related to the macroscopic shear modulus $G$ of the 
MT, which has been experimentally measured as 
$G \simeq 2 {\rm pN}/{\rm nm}^2= 2 {\rm MPa}$ (at $T=37^\circ {\rm C}$)
\cite{Kis2008}. The two-dimensional 
 shear modulus of the MT lattice is $G_{2D} = G h$, where 
$h\simeq 4{\rm nm}$ is the thickness of the 
MT lattice sheet, which is given by the size of a tubulin monomer. 
The shear energy per dimer is $e_{\rm sh} = h d r G \alpha^2/2$ 
with the distance 
$r = 2\pi R/13 \simeq 5.1{\rm nm}$  between neighboring protofilament
centers 
(where $R=(R_o+R_i)/2 = 10.5{\rm nm}$ is the mean of outer and inner
tube radius). 
In order to calculate the shear  energy of the MT lattice properly,
we have to take into account the helical pitch of 3/2 dimers 
per turn, which gives rise
to a shift $z_0 \simeq 0.92 {\rm nm}$ between neighboring protofilaments.
The rest length of lateral bond springs becomes 
$l_{0,\rm hel}= (l_0^2+z_0^2)^{1/2} \simeq 1.46 {\rm nm}$.
Shearing of the MT lattice by a small  angle $\alpha$ gives rise
to elongation of the lateral springs from their rest length 
$l_{0,\rm hel}$ by $\Delta l = \Delta z z_0/l_{0,\rm hel}$, where 
$\Delta z = \alpha r$  is the relative displacement of 
protofilaments induced by shearing, see figure \ref{fig:shear}. 
Therefore, the shear energy per dimer 
can also be written as 
\begin{equation*}
   e_{\rm sh} = \frac{1}{2} k \Delta l^2 = 
   \frac{1}{2} k \frac{z_0^2r^2}{l_{0,\rm hel}^2} \alpha^2.
\end{equation*}
Comparison with $e_{\rm sh} = h d r G \alpha^2/2$  gives 
an estimate 
\begin{equation}
    k = \frac{h d l_{0,\rm hel}^2}{z_0^2r} G \simeq 7.69 k_BT\, {\rm nm}^{-2}.
\label{eq:k}
\end{equation}

In order to estimate the bending rigidity $\kappa$ 
of single tubulin dimers,
we can use existing MD simulation results
on the distribution of protofilament curvatures.
In \cite{Grafmuller2011},
thermal fluctuations of the curvature radius 
of a protofilament consisting of three   GTP- or GDP-tubulin dimers
have been investigated by MD simulations and a distribution 
of protofilament curvature radii has been measured. 
A single protofilament with 
$N_p=3$ coupled dimers will behave as a short semiflexible polymer 
of length $N_pd$.
Each of the $N_p-1$ $\alpha\beta$-junctions
 with bending angle $\Delta \theta$ 
contributes a bending energy
$e_{\rm bend,j} =   \frac{1}{2}  \kappa \Delta \theta^2 
= \frac{1}{2}  \kappa d^2 c_j^2$, where $c=\Delta \theta/d$ is the local 
curvature. 
The mean curvature $c_m= \frac{1}{N_p-1}\sum_{j=1}^{N_p-1} c_j$ 
of the  protofilament  exhibits Gaussian fluctuations with 
$\langle c_m^2 \rangle = \langle c_j^2 \rangle = \kappa d^2/k_BT$ and, 
thus, is distributed according to 
$p_c(c_m) \propto \exp( -\kappa d^2c_m^2/2k_BT)$.
Then we can calculate the 
 corresponding distribution of mean curvature radii
$R=1/c_m$,
\begin{equation}
   p(R) = p_c(c_m(R)) \frac{1}{R^2} \propto \frac{1}{R^2} 
     \exp\left( -\frac{ d^2 \kappa}{2k_BT} \frac{1}{R^2} \right),
\end{equation}
and find a  maximum at $R^2_{\rm max} = d^2\kappa/2k_BT$. 
In the  MD simulations in \cite{Grafmuller2011} 
a most probable radius $R_{\rm max} \simeq 20{\rm nm}$ has been 
measured which leads to an estimate
\begin{equation}
    \kappa = 2R^2_{\rm max}k_BT/d^2\simeq 12.5k_BT
\label{eq:kappa}
\end{equation}
for the dimer bending rigidity $\kappa$.

In \cite{Sept2010,Grafmuller2011}, the 
 dimer bending rigidity $\kappa$ was determined 
by applying equipartition to the
 mean square thermal fluctuations of bending angles 
as measured in the MD simulations
of  protofilaments containing three dimers.
This approach gives much  higher values 
$\kappa = 1.25 \times 10^4 {\rm pN}\,{\rm nm} \sim  3000 k_BT$ \cite{Sept2010}
and $\kappa \sim 300-400 k_BT$ \cite{Grafmuller2011}, which differ from 
each other and our above estimate (\ref{eq:kappa}).
Currently we have no explanation for these discrepancies.

\begin{figure}
  \includegraphics[width=0.7\textwidth]{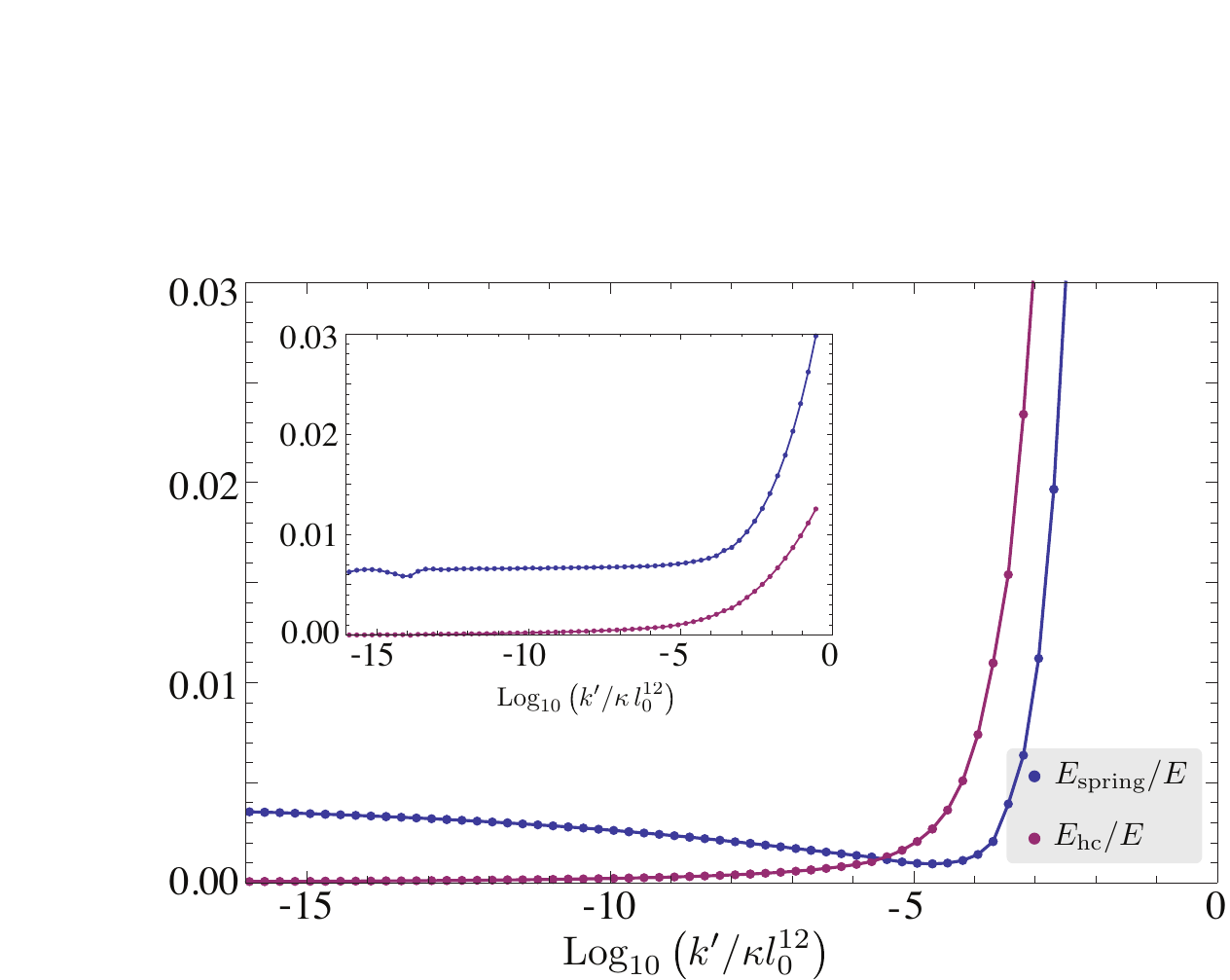}
  \caption{The fraction of hard-core energy $E_{\rm hc}/E$   of the 
total energy 
  in comparison to the  fraction of spring energy $E_{\rm hc}/E$ as a function 
of the dimensionless hard core parameter $\bar k' = k'/\kappa l_0^{12}$ for 
two different values $k/\kappa=0.5 {\rm nm}^{-2}$  and 
   $k/\kappa=0.005  {\rm nm}^{-2}$ (inset)  corresponding to strong 
and weak lateral bond springs. 
}
\label{fig:Ehc}
\end{figure}

We also have to fix the value of the 
hard core parameter $k'$. Because the hard core repulsion only serves 
as auxiliary interaction in order to avoid an unphysical overlapping 
of neighboring dimers, we want to use values for $k'$ which 
do not influence the MT equilibrium configuration 
appreciably.
Therefore, we want to choose $k'$ such that the hard core 
interaction energy $E_{\rm hc}$ remains much smaller 
than the  the bending energy $E_{\rm bend}$ 
and  the spring energy $E_{\rm spring}$.
In figure \ref{fig:Ehc},
we show the relative contributions of both energies to the 
total energy. For values  $k'/\kappa l_0^{12} < 10^{-5}$ 
the hard core interaction has negligible influence on the 
MT equilibrium configuration. We therefore 
use $k'/\kappa =10^{-6} {\rm nm}^{12}$ 
 in the following.

\begin{figure}
 \includegraphics[width=0.5 \textwidth]{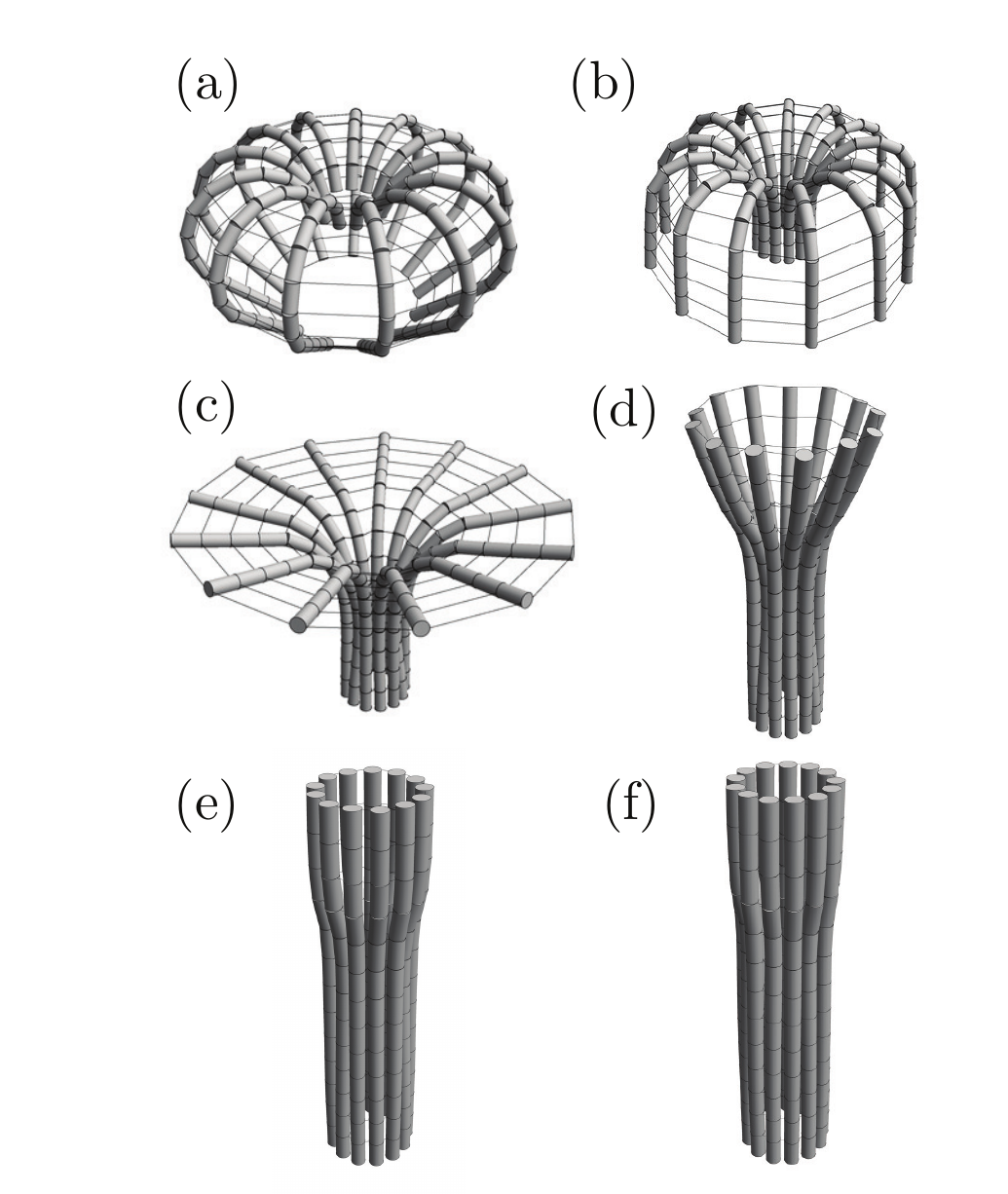}
 \caption{Mechanical equilibrium configuration of MTs for 
  $k/\kappa=$  (a) $10^{-5}$, (b)
  $1.8\times 10^{-4}$, (c) $2\times 10^{-4}$, (d) $10^{-3}$, (e) $10^{-2}$,
  and  (e) $10^{-1}$ ${\rm nm}^{-2}$
  ($k'/\kappa =10^{-6} {\rm nm}^{12}$).
  MTs are stabilized by a 3 layer GTP-cap ($m_1=10$, $m_2=3$).}
 \label{fig:smallk}
\end{figure}

Finally, 
the value of  $k/\kappa$ is constrained by the requirement
that the equilibrium MT forms a tubular structure. 
We find that for weak lateral springs, i.e., 
too small  values of $k/\kappa$ the 
stabilizing effect of a GTP-cap is lost and 
MTs spontaneously 
acquire a strongly bent shape similar to the 
 ram's horn configuration of depolymerizing catastrophic MTs.
We studied this effect systematically by calculating 
MT configurations of minimal mechanical energy for 
$k/\kappa$ in the range $k/\kappa = 10^{-5} {\rm nm}^{-2} ... 1 {\rm nm}^{-2}$,
see figure \ref{fig:smallk}.
Because such forms have not been observed experimentally,
we conclude that a reasonable lower bound  for $k/\kappa$ is 
\begin{equation}
   k/\kappa \ge 0.001 {\rm nm}^{-2}. 
\label{eq:klowerbound}
\end{equation}
Note that with our above estimates (\ref{eq:k}) for $k$ and 
(\ref{eq:kappa}) for $\kappa$ we find 
$k/\kappa \simeq 0.61 {\rm nm}^{-2}$, which is far above this bound. 
The much larger values of $\kappa$ that have been 
obtained in \cite{Sept2010,Grafmuller2011} as
discussed above  would lead to a ratio $k/\kappa$ close to 
 the lower bound 
(\ref{eq:klowerbound}). 

Because of the considerable uncertainty in the estimates 
for $k$ and $\kappa$, we will investigate values 
 $k/\kappa=0.005 {\rm nm}^{-2}$  close to  the lower bound as an example 
for  weak lateral springs and $k/\kappa=0.5  {\rm nm}^{-2}$
as an example for strong lateral springs in the following.

%%%%%%%%%%%%%%%%%%%%%%%%%%%%%
\subsection{Discussion of alternative models}

In this section, we want to discuss alternative models.
In a conceptually simpler model we could  put single interaction 
points for the lateral bond springs at the centers  of the tubulin dimers.
We investigated such simpler
 models and found that equilibrium configurations
exhibit ``wrinkled'' tubes (wrinkles forming along the 
protofilament axis) for sufficiently long MTs.
Wrinkling can not be prevented  by 
additional hard core repulsion terms. 
Therefore, such models do not agree with the experimental 
observations of tubular MTs.

Our model resembles other mechanical models proposed in
 \cite{Molodtsov2005,VanBuren2005,Molodtsov2005b}. 
The mechanical model  proposed by  Molodtsov {\it et al.} 
\cite{Molodtsov2005,Molodtsov2005b}
 also contains a  bending energy and lateral 
bonds, which are modeled as harmonic springs around their 
equilibrium length. In \cite{Molodtsov2005,Molodtsov2005b},
  four interaction points and two lateral 
springs are introduced per dimer and the helical structure is taken into 
account in contrast to our model. 
Moreover, only inter-dimer bending is considered. 
Molodtsov {\it et al.} use much larger values 
$k \sim 150 k_BT\,{\rm nm}^{-2}$ than our 
estimate $k\simeq  7.69 k_BT\,{\rm nm}^{-2}$, see (\ref{eq:k}).
Their value is obtained from an estimate of the activation energy 
for bond rupture, assuming a certain one-parameter form 
of the lateral bond potential, which tightly  couples the parameter for 
bond elasticity to the  bond rupture forces. 
We rather used values for the bond elasticity confirming 
with the macroscopic shear modulus of a protofilament sheet 
following \cite{Sept2010} and consider the bond rupture 
force as an {\it independent} parameter (see section \ref{sec:rupture} below). 
Molodtsov  {\it et al.} then also 
estimate much higher  values of $\kappa$ by applying 
a similar constraint for $k/\kappa$ as we obtained 
above (\ref{eq:klowerbound}), violation of 
which gives rise to strongly outward bending cap configurations
even in the presence of a GTP-cap. 
In contrast, we tried to obtain an estimate for $\kappa$ 
from microscopic MD simulation data. 

The mechanical model  proposed by Van Buren {\it et al.} 
\cite{VanBuren2002,VanBuren2005}
 also contains a bending energy and lateral 
bonds, which are modeled as harmonic springs of
zero rest length. In addition, the model also includes
longitudinal harmonic bond springs to model protofilament 
stretching and torsion elasticity, i.e., an elastic energy 
for the angles $\phi_j^{(i)}$, which are fixed to 
$\phi_j^{(i)} = \phi^{(i)}= 2\pi  i/13$ in our model.

Finally, we want to point out that the MT bending rigidity $\kappa_{\rm MT}$
and, thus, its persistence length, is not directly related 
to the dimer bending rigidity $\kappa$ in mechanical models. 
It is mainly related to the stiffness of longitudinal bonds between
neighboring dimers on the same protofilament, which are stretched 
and compressed in bending deformations of the tube.  
 Longitudinal bond stiffnesses are 
absent in our model as we do not consider shape fluctuations
of the whole tube but  are included in the 
models used in Refs.\ \cite{Molodtsov2005,VanBuren2005,Molodtsov2005b}
and have also been discussed in Ref.\ \cite{Pampaloni}.
Longitudinal bonds  are found to be much stiffer than lateral bonds. 
In Ref.\  \cite{Pampaloni} it has
been shown that the MT persistence length 
becomes length dependent as it contains both  bending 
 contributions related to longitudinal bonds and  shear 
contributions related to the lateral bonds. 
We do not expect the longitudinal bond stiffness, 
 the lateral bond stiffness $k$ or 
the dimer bending rigidity $\kappa$ themselves to be length dependent.

%%%%%%%%%%%%%%%%%%%%%%%%%%%%%%%%%%%%%%%%%%%
\subsection{Coupling of hydrolysis and microtubule mechanics}

In the allosteric model,
hydrolysis of GTP-tubulin dimers to GDP-tubulin dimers within the 
MT leads to bending of the dimer, i.e., 
a change in the equilibrium dimer angle 
$\theta_0$. This results in mechanical forces and torques in the MT lattice, 
which are transmitted by the harmonic binding forces between 
protofilaments and which drive the MT
lattice to a new equilibrium state.

We can view  $\theta_0$ as the reaction coordinate of hydrolysis such 
that, in the absence of mechanical forces, hydrolysis of single dimers
is characterized by a  free energy profile 
$F_h(\theta_0)$ 
with a free energy minimum corresponding to a GTP-dimer at $\theta_0=0^\circ$
and a second minimum corresponding to  the GDP-dimer state at
$\theta_0=22^\circ$. Because of the free energy $\Delta G_{\rm GTP} \simeq 5k_BT$
released by hydrolysis of GTP within the MT lattice \cite{Desai1997}, 
the second minimum is lower by at most 
$F_h(0^\circ)-F_h(22^\circ)< \Delta G_{\rm GTP} \simeq 5k_BT$.

Because GTP hydrolysis can generate mechanical forces within the
MT,  mechanical forces also influence the hydrolysis 
rate of GTP-dimers. 
This influence can be quantified using Bell theory \cite{Bell}.
For each equilibrium angle $\theta_0\equiv \theta_{0,j}^{(i)}$ 
of a given dimer 
we can also calculate the mechanical 
total equilibrium energy $E(\theta_0)$ of the MT lattice.
The relevant free energy profile for hydrolysis, that takes into 
account the mechanics of the surrounding MT lattice, is 
$F_h(\theta_0) + E(\theta_0)$.
Using $\theta_0$ as a reaction coordinate there will be 
a rate limiting free energy barrier $\Delta F_h$ for an 
intermediate $0^\circ < \theta_{0,\rm max} < 22^\circ$. 
The height and the exact angle $\theta_{0,\rm max}$
of this energy barrier will also be modified by 
the corresponding mechanical contribution
to $\Delta F_h + \Delta E$. If we neglect a possible 
shift of the barrier angle, we have
$\Delta E = E(\theta_{0,\rm max})-E(0^\circ)$.
We  have currently no information on the detailed free energy profile
$F_h(\theta_0)$  of 
the hydrolysis reaction; we only have information on 
 the hydrolysis rates themselves, which should be  related to the 
barrier height $\Delta F_h$ via the Arrhenius law.
To proceed, we  simply 
assume  $\theta_{0,\rm max} = 11^\circ$ in the following. 
One important point regarding the influence of the mechanics 
of the MT lattice onto the hydrolysis rates is the following:
{\it differences} between hydrolysis rates at different 
sites in the MT lattice are not governed by $\Delta F_h$,
which is identical for all MT lattices sites, but rather by the 
mechanical contribution $\Delta E$ to the barrier, which we 
want to estimate now.

We know that the total mechanical energy 
is a   quadratic function of 
$\theta_0$ via the bending energy contribution in 
 (\ref{eq:long_int}). In order to determine $\Delta E$ 
for hydrolysis we have to  increase $\theta_0$ from $0^\circ$ to 
$\theta_{0,\rm max}$ to obtain the saddle point value $\Delta E$.
Because we assume that a single hydrolysis reaction is faster 
than mechanical relaxation of the MT lattice, we 
increase $\theta_0$ for 
{\it fixed} values of the configuration angles $\theta_j^{(i)}$. 
We assume an approximately  linear dependence
$E(\theta_{0})-E(0^\circ) \approx (E(\delta)-E(0^\circ)) \theta_{0}/\delta$ 
for a small angle $\delta$, and  estimate $\Delta E$ numerically
 using $\delta = 1^\circ$ and assuming $\theta_{0,\rm max}= 11^{\circ}$
for the position of the barrier. 
This results in the estimate
\begin{equation}
  \Delta E  = E(11^\circ) -  E(0^\circ)\simeq  11 (E(1^\circ)-E(0^\circ))
\label{eq:DeltaE}
\end{equation}
(calculated for fixed values of all configuration angles $\theta_j^{(i)}$).
According to its definition, $\Delta E$ will be small (eventually 
even negative) for hydrolysis of a given dimer if there are 
mechanical forces in the MT lattice 
 that pull the dimer into its bent GDP-configuration. 

This mechanical shift  of the free energy barrier for 
the hydrolysis reaction  will modulate the hydrolysis rate  $r_h$
of a dimer according to 
\begin{equation}
  r_h(\Delta E) = r_h(0) \exp(-\Delta E/k_BT)
\label{eq:rDeltaE}
\end{equation}
following Bell  \cite{Bell}.
In particular, this leads to {\it site-dependent} hydrolysis rates, 
which depend on the position of the dimer in the MT lattice via 
the mechanical forces acting on it at that position. 
We conclude that the interaction via the mechanics of the MT lattice 
can give rise to possible correlation effects in the hydrolysis 
dynamics, which have not been taken into account before.

Our  simulation algorithm for the 
most probable hydrolysis pathway (see section \ref{sec:hydrolysis_path} below) 
 will be  based on two assumptions
regarding the time scales for hydrolysis and mechanical relaxation:
(i) a single hydrolysis reaction is fast compared to mechanical 
  relaxation, which was the basis for calculating 
 the modulated hydrolysis rate (\ref{eq:rDeltaE}),
and (ii) mechanical relaxation is faster than the time {\it between} 
  successive hydrolysis events (set by the hydrolysis rate $r_h^{-1}$) such 
that we relax the MT lattice mechanically 
between successive hydrolysis events.

%%%%%%%%%%%%%%%%%%%%%%%%%%%%%%%
\subsection{Hydrolysis and mechanical model parameters}

Within the MT GTP-dimers are hydrolyzed into GDP-dimers. 
The backward reaction is not observed. This puts a 
constraint on the mechanical model parameters because 
it implies that the total mechanical 
energy difference during hydrolysis of a certain dimer,
$\Delta E_h = E(22^\circ)- E(0^\circ)$, has to 
be much smaller than 
the chemical energy $\Delta G_{\rm GTP}\simeq 
5k_BT$ released by GTP hydrolysis in the absence of 
any mechanical forces \cite{Desai1997}, 
$\Delta E_h \ll \Delta G_{\rm GTP}$.

We consider the situation where the MT is still stabilized by the 
GTP-cap to a tubular configuration such that all angles $\theta_j^{(i)}$ 
are small. Then, hydrolysis of a single dimer (in layer $j$ and 
protofilament $i$) increases its  rest angle 
and strains the surrounding MT lattice.
If the lateral interaction springs are sufficiently strong 
to stabilize a tubular MT configuration, the change $\Delta E_{\rm spring}$ 
in the lateral spring energies by this hydrolysis can be neglected 
as compared to  the change $\Delta E_{\rm bend}$ in bending 
energy  such that the total mechanical 
energy increase during hydrolysis is approximated by 
\begin{eqnarray}
\Delta E_h \approx  \Delta E_{\rm bend} 
   &=& \frac{\kappa}{2} \left( \theta_j^{(i)}-\theta_{j-1}^{(i)}
              - 22^\circ \right)^2 - 
       \frac{\kappa}{2} \left( \theta_j^{(i)}-\theta_{j-1}^{(i)}\right)^2
      \nonumber\\
   &\approx &  \frac{\kappa}{2} (22^\circ)^2.
\label{eq:Eh}
\end{eqnarray}
Then, the condition $\Delta E_h \ll \Delta G_{\rm GTP}\simeq 5k_BT$
 implies an upper bound on the dimer bending rigidity $\kappa$, 
\begin{equation}
   \kappa  \ll 2 \Delta G_{\rm GTP} \frac{1}{(22^\circ)^2} \simeq 
       68 k_BT\,{\rm rad}^{-2}.
  \label{eq:kappa_upperbound}
\end{equation}
Our above estimate (\ref{eq:kappa}), $\kappa \simeq 12.5 k_BT$, from 
analyzing MD simulation results for protofilament curvature 
radii distributions obeys the constraint (\ref{eq:kappa_upperbound}).
The other estimates  $\kappa\sim 3000k_BT$ from \cite{Sept2010}
and $\kappa \sim 300 k_BT$ from \cite{Grafmuller2011},
however, violate this constraint.  If $\kappa$-values violating 
the upper bound (\ref{eq:kappa_upperbound}) are confirmed 
experimentally in the future, this can be a hint that the 
allosteric model itself, i.e., the assumption that 
GTP hydrolysis changes the tubulin dimer angle,  is inconsistent.

Within the allowed  upper 
bound $\Delta E_h \le \Delta G_{\rm GTP}$,
we can distinguish two limits
for the hydrolysis rates: (i) if $\Delta E \ll k_BT$, the influence 
of the mechanical shift $\Delta E$ of the free energy barrier for 
the hydrolysis reaction is  small, and we expect purely ``chemical'' 
models as defined in the introduction to be essentially correct. 
(ii) If $\Delta E \gg k_BT$, on the other hand, 
the mechanical shift $\Delta E$ dominates the hydrolysis rates 
according to (\ref{eq:rDeltaE}) and we expect a strong interplay
between MT mechanics and hydrolysis. 

In  limit (ii), the dimer to be hydrolyzed next is mainly determined by 
 the  mechanical shifts $\Delta E$ of the free energy barrier for 
the hydrolysis reaction; the GTP-dimer with the smallest $\Delta E$ 
will be hydrolyzed next, eventually under additional 
restrictions depending on whether we consider a vectorial or random 
hydrolysis mechanism. 
Only if several GTP-dimers have a similar $\Delta E$, the next 
hydrolysis event will be stochastic among these dimers. 
Therefore, 
 the order of hydrolysis of GTP-dimers within the MT
lattice (the hydrolysis pathway) 
is mainly determined by the hydrolysis mechanism 
(vectorial or random) {\it and} the mechanical energies
$\Delta E$ and exhibits much less stochasticity as compared 
to limit (i), as discussed below in the results section.  

We will now explore under which condition we can expect 
such mechanically dominated hydrolysis order, i.e., 
under which conditions $\Delta E \ge k_BT$ holds. 
The mechanical shift $\Delta E$ of the energy barrier for the 
hydrolysis reaction 
 should be smaller than total 
mechanical energy change $\Delta E_h$ during hydrolysis,
$\Delta E \le \Delta E_h$ 
resulting in a condition  $\Delta E_h \ge \Delta E \ge k_BT$ or 
(using (\ref{eq:Eh}))
\begin{equation}
   \kappa  \ge  2 k_BT \frac{1}{(22^\circ)^2} \simeq 
       14 k_BT\,{\rm rad}^{-2}.
  \label{eq:kappa_lowerbound}
\end{equation}
For smaller  values of $\kappa$, we expect to find hydrolysis rates,
which are essentially independent of mechanical forces developing 
in the MT lattice.

Our above estimate (\ref{eq:kappa}), $\kappa \simeq 12.5 k_BT$, from 
analyzing MD simulation results for protofilament curvature 
radii distributions violates 
 condition (\ref{eq:kappa_lowerbound}) only weakly.
The other estimates for $\kappa$ from \cite{Sept2010,Grafmuller2011},
however, give much higher values of $\kappa$ supporting mechanically
dominated hydrolysis rates.

%%%%%%%%%%%%%%%%%%%%%%%%%%%%
\section{Results}

We use the mechanical model introduced above to investigate 
how mechanical forces influence and direct hydrolysis 
pathways and how catastrophe events emerge if lateral 
bond rupture is included.

%%%%%%%%%%%%%%%%%%%%%%%%%%%%%%%%
\subsection{Hydrolysis Pathways}
\label{sec:hydrolysis_path}

We explore results for a mechanically 
dominated hydrolysis rate where the 
mechanical shift  of the free energy barrier for 
the hydrolysis reaction is  much larger 
than the thermal energy $1k_BT$. 
Therefore, our results   will apply if 
both bounds (\ref{eq:kappa_upperbound}) and (\ref{eq:kappa_lowerbound}) 
are met, i.e., in a range 
$14 k_BT \le \kappa \ll 68 k_BT$ of dimer bending rigidities. 
This regime is conceptually interesting because 
correlation effects in the hydrolysis dynamics
introduced via the mechanics of the MT lattice 
become maximal. 
Moreover, 
this  parameter range for the  tubulin dimer bending rigidity 
$\kappa$ cannot be ruled out at present 
because different estimates for $\kappa$ are deviating 
and not reliable. 
Our above estimate $\kappa\simeq 12.5 k_BT$, see 
(\ref{eq:kappa}), is close 
to the  lower boundary of the considered $\kappa$-range.

If hydrolysis is mechanically dominated ($\Delta E \gg k_BT$), 
the hydrolysis pathways, i.e., 
the order of hydrolysis  of dimers
within the MT, is mainly determined by the mechanical 
 shift $\Delta E$ of the energy barrier for the 
hydrolysis reaction: the  dimer to be hydrolyzed  
next with highest probability
is the GTP-dimer with the smallest $\Delta E$ among all 
GTP-dimers accessible by the hydrolysis mechanism.
For random hydrolysis all GTP-dimers are accessible, for 
vectorial hydrolysis only the GTP-dimers at the cap boundary.
Because the dimensionless
energies $\Delta E/\kappa$ only depend on the parameter 
$k/\kappa$, the hydrolysis pathway is entirely determined by this 
mechanical parameter for both mechanisms. 
As a result of mechanically dominated hydrolysis, 
the choice of a certain hydrolysis pathway 
becomes much less stochastic: from the large number of 
chemically possible pathways only relatively few have 
considerable statistical weight if the influence of 
mechanics is dominant for $\Delta E \gg k_BT$.
Therefore, the concept of a {\it most probable hydrolysis 
pathway}, which is the 
hydrolysis pathway with the highest statistical weight,
 is reasonable in this limit.

In the simulation, we determine the most probable hydrolysis pathway 
by the following algorithm:
We relax the mechanical forces in the MT lattice by energy minimization
for a given hydrolysis state. Next, we calculate the mechanical 
 energy shifts $\Delta E$  for each GTP-dimer which can be hydrolyzed 
according to the assumed hydrolysis mechanism (random or vectorial). 
We calculate $\Delta E$ according to (\ref{eq:DeltaE}). 
Then, out of these GTP-dimers, we choose the one with the minimal
$\Delta E$ and, thus, the highest hydrolysis rate according to 
(\ref{eq:rDeltaE}) to be hydrolyzed next. Then, we  
relax  the MT lattice again mechanically, and so on.

The condition $\Delta E \gg k_BT$ is necessary but not sufficient 
to select a  hydrolysis pathway  uniquely, i.e., 
render the hydrolysis deterministic:
if several GTP-dimers have a similar $\Delta E$, the next 
hydrolysis event will be essentially stochastic among these dimers. 
The hydrolysis pathway becomes deterministic only if 
also  {\it differences} $\Delta \Delta E$ 
between the $\Delta E$  for different GTP-dimers
within the MT are  much larger than the thermal 
energy, i.e.,  $\Delta \Delta E \gg k_BT$.

As discussed in the introduction, 
the correct  hydrolysis mechanism is not exactly known. 
Therefore we consider both vectorial and random hydrolysis separately
in the following.
We address the question how the mechanical forces direct the 
hydrolysis pathway for both mechanisms and as a function of 
the parameter $k/\kappa$. We will consider two exemplary values:
 $k/\kappa=0.005 {\rm nm}^{-2}$ close to  the lower bound
 (\ref{eq:klowerbound}) as an example 
for  weak lateral springs and $k/\kappa=0.5  {\rm nm}^{-2}$
as an example for strong lateral springs in the following.

In order to isolate effects of MT mechanics onto the 
hydrolysis pathway,
we  ignore the polymerization dynamics and consider 
hydrolysis in MTs of fixed length 
with a stabilizing GTP-cap consisting of 
3 layers ($m_1=20$ and $m_2=3$).

%%%%%%%%%%%%%%%%%%%%%%
\subsubsection{Random  hydrolysis.}

For random hydrolysis the chemical hydrolysis rate $r_h(0)$ 
of a given GTP-dimer 
is independent of the hydrolysis state of its neighbors, 
see (\ref{eq:rDeltaE}). 
Therefore, all GTP-dimers in the cap can be hydrolyzed 
with equal probability if mechanics is neglected.

\begin{figure}
 \includegraphics[width=0.8\textwidth]{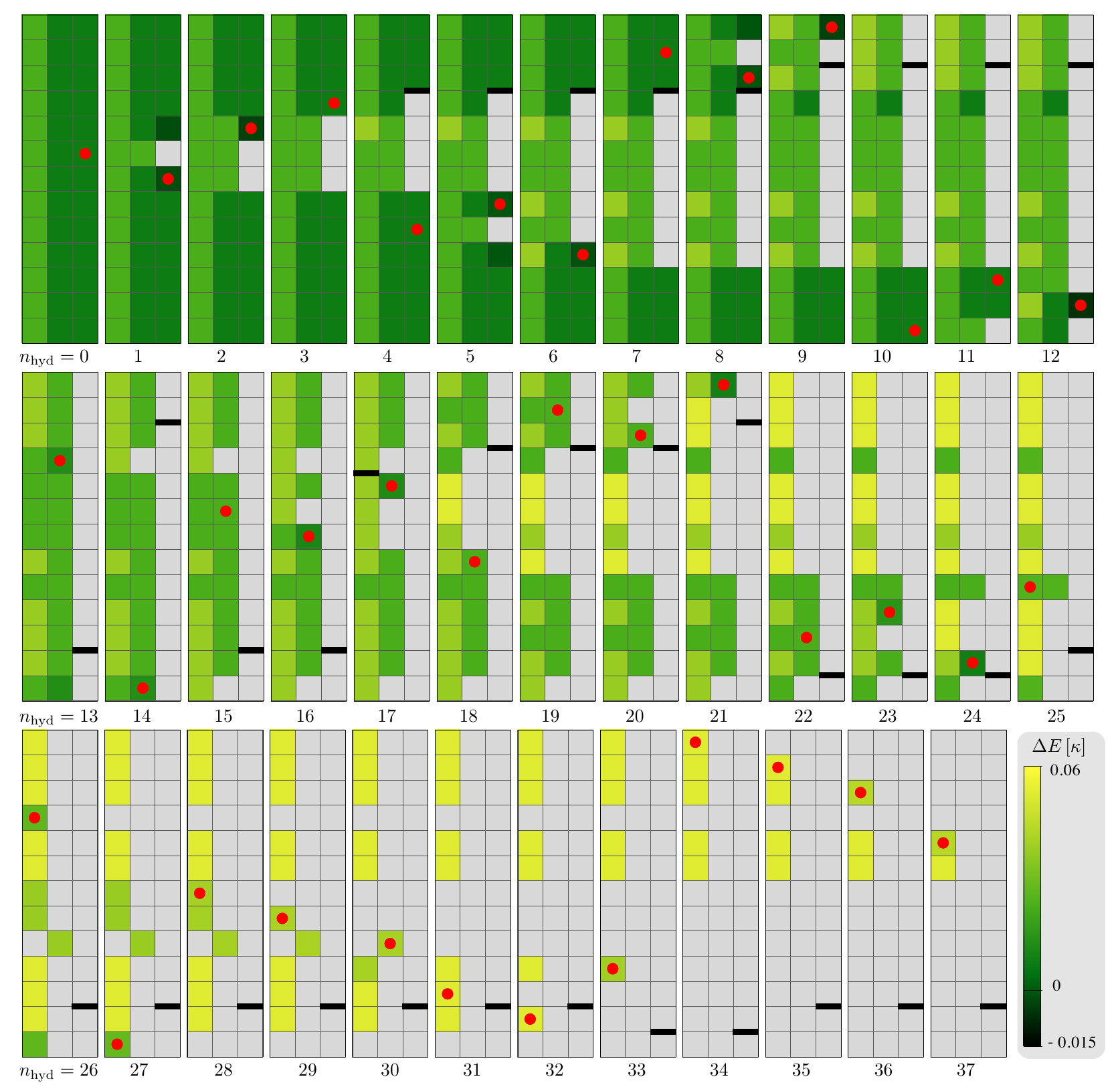}
\caption{
  Random hydrolysis:
   Most probable hydrolysis pathway of a  
   MT with a 3 layer GTP-cap with 
   strong lateral bond springs ($k/\kappa=0.5  {\rm nm}^{-2}$).
   Each 13x3 rectangle in the sequence of 38 hydrolysis steps 
   shows the hydrolysis 
  state of the 3 layer GTP-cap (MT plus end on the right side):
   squares symbolize dimers; grey squares represent 
    hydrolyzed GDP-dimers; green squares represent GTP-dimers
   color-coded for their respective $\Delta E/\kappa$.
   The red dot marks the dimer to be hydrolyzed next.
  The thick black vertical line marks the lateral bond 
  under maximal force, see figure \ref{fig:kraft_random}a.
}
 \label{fig:HP_random_05}
\end{figure}

\begin{figure}
 \includegraphics[width=0.8\textwidth]{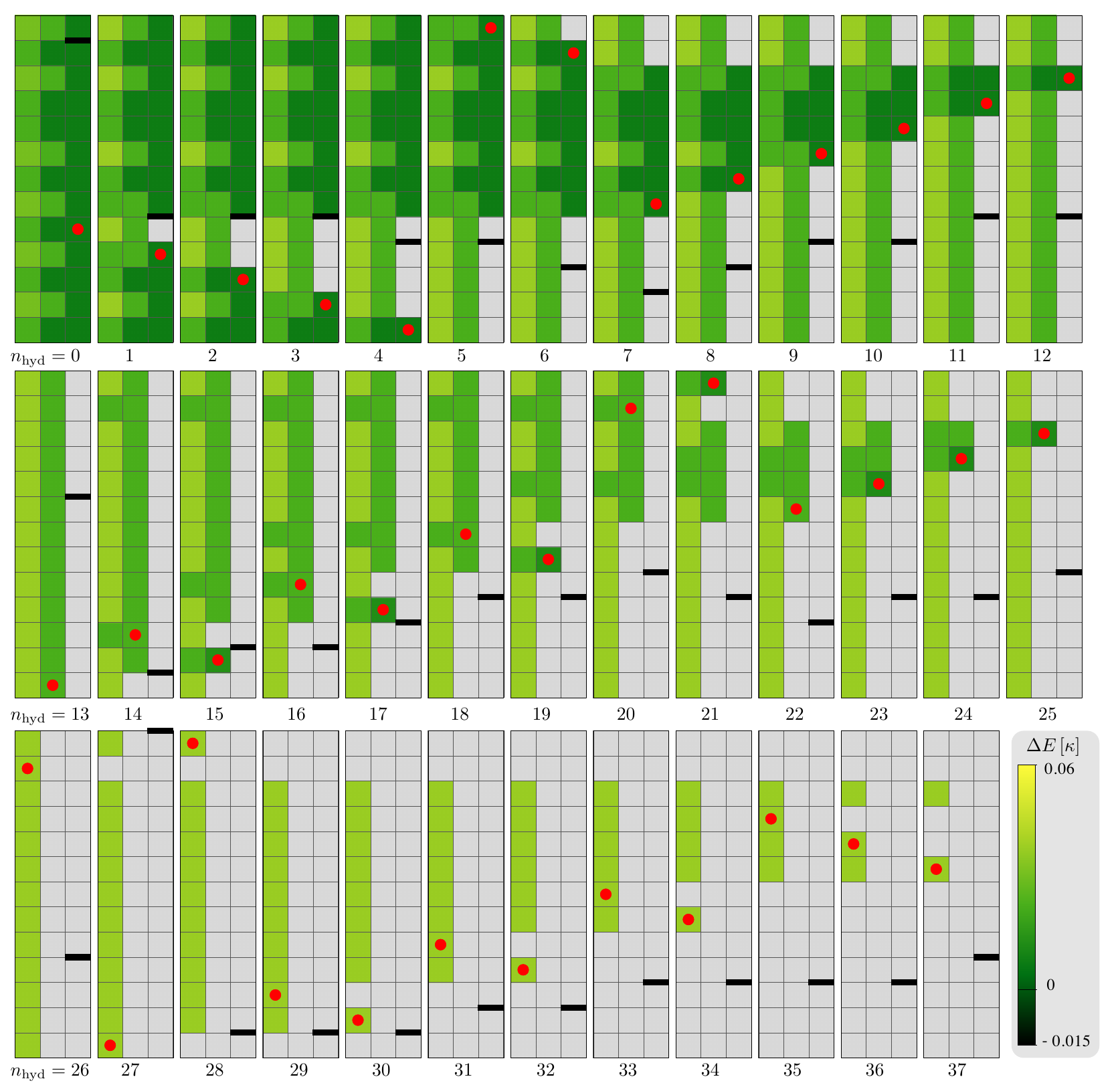}
\caption{
   Random hydrolysis:
  Most probable hydrolysis pathway of a  
   MT with a 3 layer GTP-cap  with 
    weak  lateral bond springs ($k/\kappa=0.005  {\rm nm}^{-2}$).
  Representation as in figure \ref{fig:HP_random_05}. 
  The thick black vertical line marks the lateral bond 
  under maximal force, see figure \ref{fig:kraft_random}b.
}
 \label{fig:HP_random_0005}
\end{figure}

For random hydrolysis in the absence of mechanical forces, all GTP-dimers 
in the cap are accessible for hydrolysis 
with equal probability, and in a MT with a GTP-cap size of 
three layers ($m_2=3$) there are $39!$ equally probable 
hydrolysis pathways. 
For mechanically dominated 
random hydrolysis, on the other hand, 
it is always  the dimer with 
smallest $\Delta E$ among all GTP-dimers, which is hydrolyzed next 
with the highest probability.
As a result there emerges a {\it most probable hydrolysis pathway},
which is shown 
in figures \ref{fig:HP_random_05}  and \ref{fig:HP_random_0005}
 for a   MT with $m_1=20$ and $m_2=3$ and  strong 
springs  ($k/\kappa=0.5  {\rm nm}^{-2}$, figure \ref{fig:HP_random_05}) 
or weak springs 
 ($k/\kappa=0.005  {\rm nm}^{-2}$, figure \ref{fig:HP_random_0005}).
These figures show the sequence of 38 hydrolysis states 
 of the 3x13 dimer cap (from $n_{\rm hyd}=0$ GDP-dimers to $n_{\rm hyd}=37$ 
GDP-dimers).
 GTP-dimers are represented
by green squares, hydrolyzed GTP-dimers  by grey squares. 

 If there are several GTP-dimers with similarly small 
 $\Delta E$, such that $\Delta \Delta E \le k_BT$, 
they can be hydrolyzed with comparable probability,
and there are other   hydrolysis pathways which are 
similarly probable.  
The values of $\Delta E$ of different GTP-dimers 
(measured in units of $\kappa$) are 
indicated  in the most probable hydrolysis pathways in 
figures  \ref{fig:HP_random_05}  
and \ref{fig:HP_random_0005} by color-coding. 
The existence of a unique, dark green square in a cap indicates 
the existence of a unique next hydrolysis spot, which is 
separated by a large $\Delta \Delta E$ from other possible 
hydrolysis spots. 
Small color differences indicate small values of $\Delta \Delta E$, 
such that the next 
hydrolysis spot  should  be chosen essentially stochastic 
among the dimers with darkest green colors. 
Then there exist other pathways competing to the ones 
shown in figures  \ref{fig:HP_random_05}  
and \ref{fig:HP_random_0005}, which occur with comparable 
probability.

For random hydrolysis,   figures \ref{fig:HP_random_05}  
and \ref{fig:HP_random_0005} show that the most probable 
hydrolysis pathways
are very similar for strong and weak lateral bond springs. 
Two features of the influence of mechanics can be deduced 
from   figures \ref{fig:HP_random_05}  
and \ref{fig:HP_random_0005}:

(i)  Hydrolysis is favored in the front  layer
  of the GTP-cap towards the plus end (see states $n_{\rm hyd}= 1, 14, 27$). 
   In the front layer, the mechanical strain 
  from the bending of the GDP-dimers behind the cap 
   gives rise to the strongest outward bending moments 
   on a dimer via outward pulling forces exerted 
  by the lateral springs. 

(ii) After hydrolysis has started in a layer, there is a 
strong preference that this layer becomes completely 
  hydrolyzed before it proceeds in the next layer.
   Together with point (i) this gives rise to a very 
  regular hydrolysis pathway if mechanics is taken into account,
   although the underlying chemical hydrolysis rule is random. 
  
(iii) There is a strong preference for GTP-dimers 
  with two lateral GDP-neighbors to be hydrolyzed (see, for example, 
    states  $n_{\rm hyd}= 12, 17, 20, 21$  in figure \ref{fig:HP_random_05}).
   These isolated straight GTP-dimers are pulled outward 
   by already hydrolyzed and, thus, bent GDP-neighbors.

%%%%%%%%%%%%%%%%%%%%%%%%
\subsubsection{Vectorial  hydrolysis.}

For vectorial hydrolysis, only dimers whose longitudinal 
neighbor (towards the minus end) 
 is already hydrolyzed into GDP-tubulin are accessible
for hydrolysis. 
Then there is a sharp interface between GDP- and GTP-dimers,
which propagates towards the plus end. For mechanically 
dominated vectorial hydrolysis, the GTP-dimer with 
the smallest $\Delta E$  among all GTP-dimers at the 
GTP-GDP interface is hydrolyzed next. 

\begin{figure}
 \includegraphics[width=0.8\textwidth]{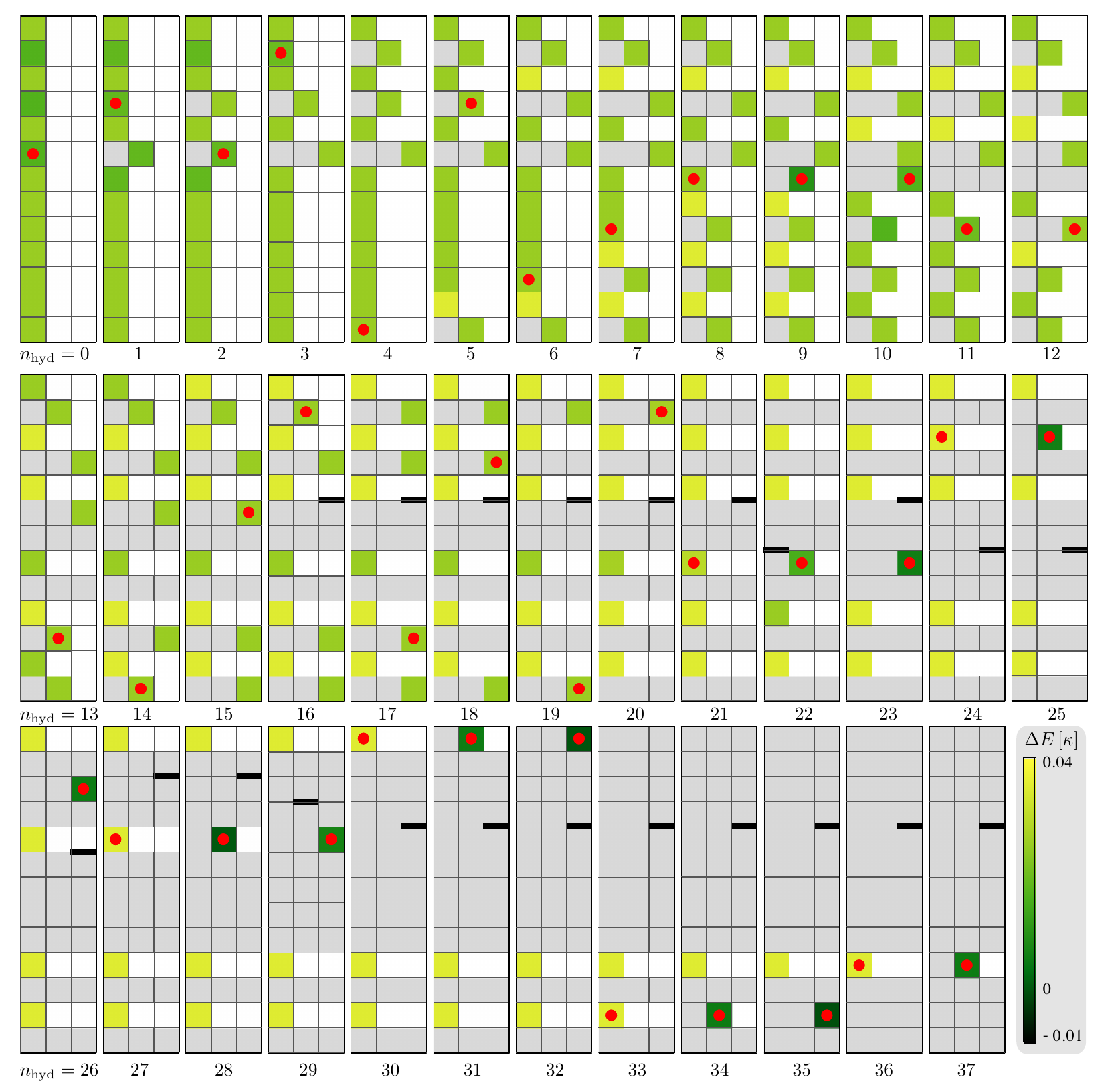}
\caption{ 
   Vectorial hydrolysis:
    Most probable hydrolysis pathway of a  
   MT with a 3 layer GTP-cap  with
   strong lateral bond springs ($k/\kappa=0.5  {\rm nm}^{-2}$).
Each 13x3 rectangle in the sequence of 38 hydrolysis steps 
   shows the hydrolysis 
  state of the 3 layer GTP-cap (MT plus end on the right side):
    squares symbolize dimers; grey squares represent 
    hydrolyzed GDP-dimers; green squares represent GTP-dimers
   at the boundary of the GTP-cap,
   color-coded for their respective $\Delta E/\kappa$.
   White squares represent GTP-dimers, which are not at the 
   cap boundary and, thus, cannot be hydrolyzed in a vectorial 
   mechanism. The red dot marks the dimer to be hydrolyzed next.
 The thick black vertical line marks the lateral bond 
  under maximal force, see figure \ref{fig:kraft_vec}a.
 }
 \label{fig:HP_vec_05}
\end{figure}

\begin{figure}
 \includegraphics[width=0.8\textwidth]{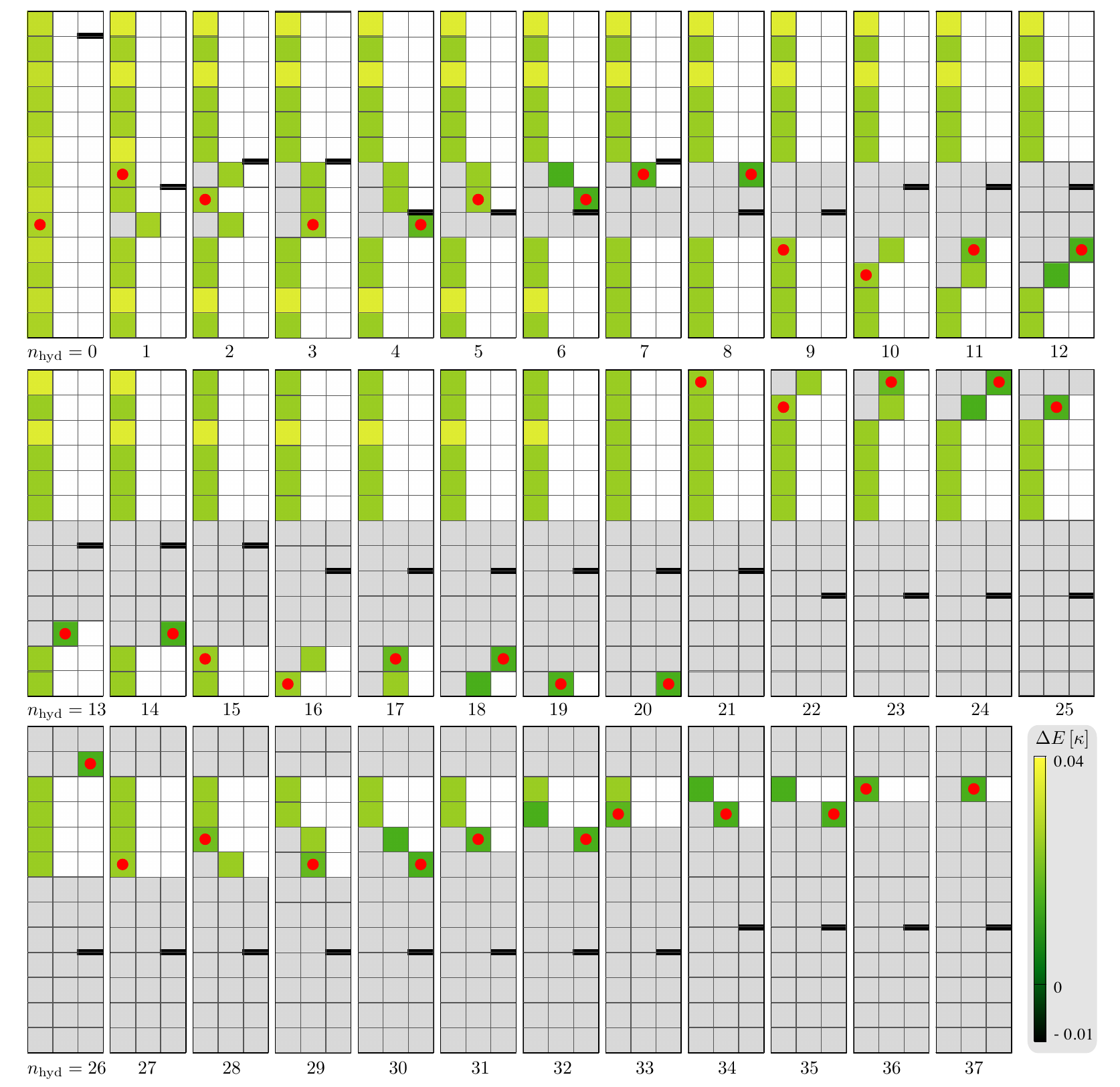}
\caption{
   Vectorial hydrolysis:
    Most probable hydrolysis pathway of a  
   MT with a 3 layer GTP-cap  with 
   weak lateral bond springs 
    ($k/\kappa=0.005  {\rm nm}^{-2}$).
  Representation as in figure \ref{fig:HP_vec_05}.
   The thick black vertical line marks the lateral bond 
  under maximal force, see figure \ref{fig:kraft_vec}b.}
 \label{fig:HP_vec_0005}
\end{figure}

Again, for mechanically dominated 
vectorial hydrolysis, there  is  a most probable hydrolysis pathway,
which is shown 
in figures \ref{fig:HP_vec_05}  and \ref{fig:HP_vec_0005}
 for a   MT with $m_1=20$ and $m_2=3$ and  strong 
springs  ($k/\kappa=0.5  {\rm nm}^{-2}$, figure \ref{fig:HP_vec_05}) 
or weak springs 
 ($k/\kappa=0.005  {\rm nm}^{-2}$, figure \ref{fig:HP_vec_0005})
in terms of the sequence of 38 hydrolysis states of the
3x13 dimer cap. For vectorial hydrolysis, there are 
also GTP-dimers which cannot be hydrolyzed because they are 
not at the GTP-GDP interface, i.e., there is no hydrolyzed 
neighboring GDP-dimer towards the minus end. 
These dimers are shown in white color. 
Only hydrolyzable GTP-dimers at the GTP-GDP interface are 
represented 
by green squares, hydrolyzed GTP-dimers  by grey squares
as before. 

We recognize marked differences as compared to the 
random hydrolysis:

(i) Because of the vectorial constraint, 
   hydrolysis  has to start in the back  layer
  of the GTP-cap towards the minus end, where the GTP-GDP
  interface is located. It cannot advance directly to the front 
   layer as for random hydrolysis. 

(ii) There is a strong tendency to continue hydrolysis 
on the same protofilament (see states $n_{\rm hyd}= 1, 2, 3$
 or $n_{\rm hyd}= 9, 10, 11$
 in figure \ref{fig:HP_vec_05}). 
  If the hydrolysis front has advanced to the next layer on 
  one protofilament, the bending forces from the hydrolyzed 
  dimers behind the front give rise to a strong outward 
  bending moment on the next GTP-dimer, which favors its
  hydrolysis. This force is transmitted via the lateral 
   bonds to neighboring 
    protofilaments in the MT lattice.

(iii) As for random hydrolysis, 
    there is a preference for GTP-dimers 
  with one or two lateral GDP-neighbors to be hydrolyzed (see, for example, 
    states $n_{\rm hyd}= 10, 22, 23, 25, 26$ in figure \ref{fig:HP_vec_05}).

 (iv) The effects (ii) and (iii) combine to a ``nucleation-like'' behavior:
   Once a nucleus of three neighboring hydrolyzed  GDP-dimers 
  has formed in a layer, the hydrolysis front tends to advance up to 
  the plus end  on one of the protofilaments 
    (states $n_{\rm hyd}= 3, 4, 5$ in figure \ref{fig:HP_vec_0005}).
   This effect is more pronounced for weak lateral bond springs.

In the next section we will see that the first advance of the hydrolysis 
front to the plus end is typically linked to a pronounced increase 
of the lateral bond forces, in particular for weak 
lateral bonds. If lateral bonds are allowed to rupture, this 
increase in forces  can initiate  catastrophe events. 
The last feature (iv) of a nucleus of three GDP-dimers that has to assemble 
before the hydrolysis front will reach the plus end could be an explanation 
of the experimental finding of two rate-limiting steps, which 
are involved in the initiation of a catastrophe \cite{Gardner2011a}.

%%%%%%%%%%%%%%%%%%%%%%%%%%%%%%%%%%%%%%%%%%%%%
\subsection{Lateral bond rupture and catastrophes}
\label{sec:rupture}

During a catastrophe, it is experimentally observed that 
the protofilaments of a MT fall apart and curl into 
``ram's horn'' conformations as a result of the $22^\circ$ equilibrium 
angle of hydrolyzed GDP-dimers. 
This implies that protofilaments separate during catastrophes, 
i.e., lateral bonds between dimers in neighboring protofilaments
will rupture. 

Bond rupture is an activated process, and the 
bond rupture rate under force, 
$r_{\rm rup}(F)  = r_{\rm rup}(0) \exp(F/F_{\rm  rup})$,
 increases exponentially  
above a characteristic rupture force  $F_{\rm rup}$
according to  Bell theory \cite{Bell}.
 In a simplified approach, we can ignore stochastic 
rupture effects and assume that all lateral bonds under forces $F>F_{\rm rup}$ 
rupture. 
One problem is that there is currently no microscopic information 
on values for the rupture force $F_{\rm rup}$. 
In \cite{Molodtsov2005b}, an activation energy for the 
rupture process was estimated from the depolymerizing velocity 
of catastrophic MTs as $E_{\rm rup} \simeq 10k_BT$. 
Together with a characteristic bond length of $\simeq 0.25{\rm nm}$ 
used in \cite{Molodtsov2005b}, one arrives at very large 
rupture forces $F_{\rm rup} \sim 160 {\rm pN}$.

%%%%%%%%%%%%%%%%%%%%%%
\subsubsection{Maximal lateral bond force.}

\begin{figure}
\includegraphics[width=0.8\textwidth]{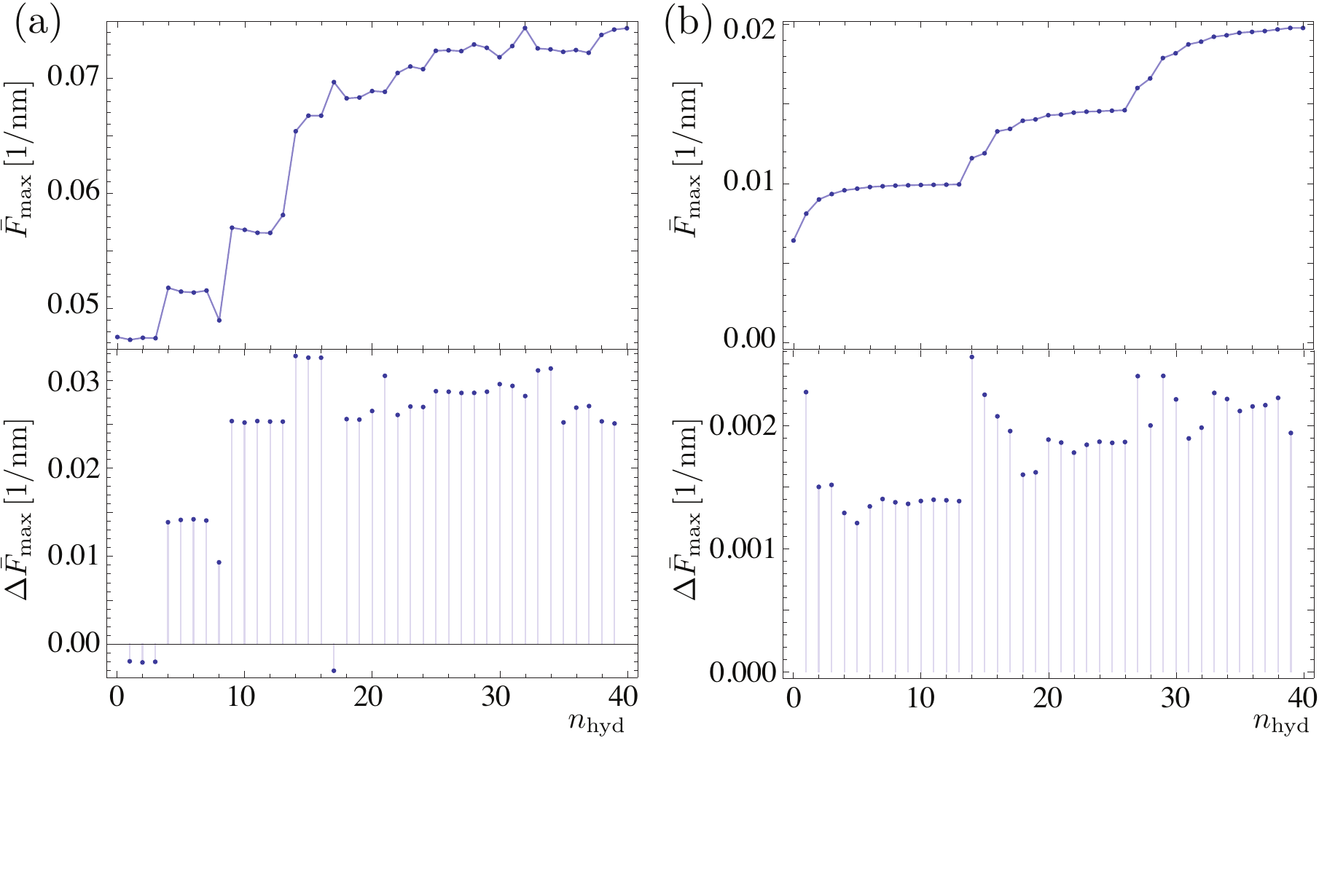}
\caption{
  Random hydrolysis:
   Maximal lateral bond force $\bar F_{\rm max} = F_{\rm max}/\kappa$
   and change $\Delta \bar F_{\rm max}$ in the maximal lateral bond force 
  after rupture of the maximally strained bond as a function 
   of the number $n_{\rm hyd}$ of hydrolyzed GTP-dimers. We start with a  
   MT with a 3 layer GTP-cap.  a) Strong lateral 
  bond springs, $k/\kappa=0.5  {\rm nm}^{-2}$,
    and b) weak lateral bond springs, $k/\kappa=0.005  {\rm nm}^{-2}$.
  The location of the lateral bond under maximal force within the GTP-cap 
  is shown in figures \ref{fig:HP_random_05} and \ref{fig:HP_random_0005}
  for strong and weak bonds, respectively, as thick black line. 
}
 \label{fig:kraft_random}
\end{figure}

\begin{figure}
 \includegraphics[width=0.8\textwidth]{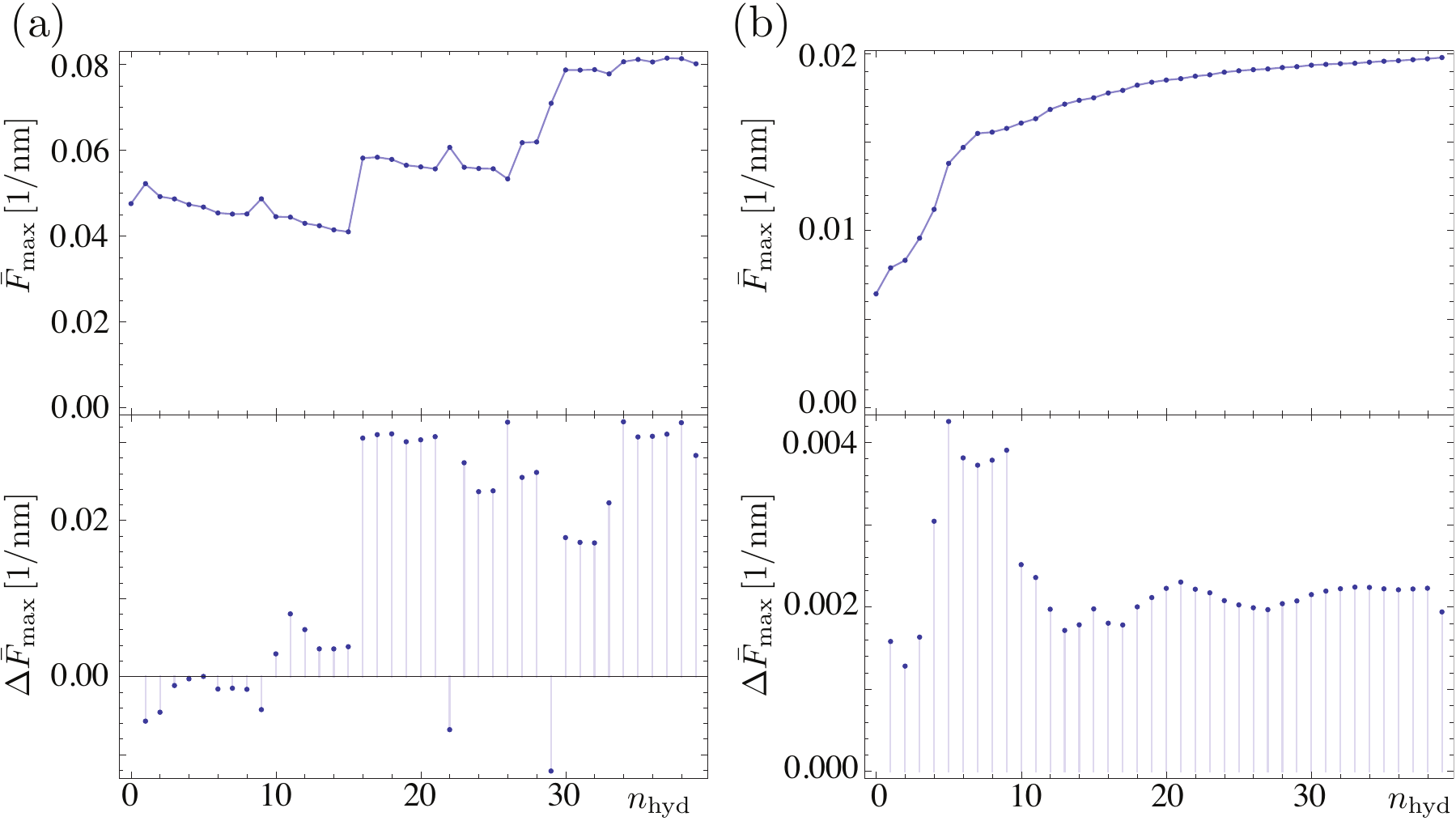}
\caption{
   Vectorial  hydrolysis:
   Maximal lateral bond force $\bar F_{\rm max} = F_{\rm max}/\kappa$
   and change $\Delta \bar F_{\rm max}$ in the maximal lateral bond force 
  after rupture of the maximally strained bond as a function 
   of the number $n_{\rm hyd}$ of hydrolyzed GTP-dimers. We start with a  
   MT with a 3 layer GTP-cap. a) Strong lateral 
  bond springs, $k/\kappa=0.5  {\rm nm}^{-2}$,
    and b) weak lateral bond springs, $k/\kappa=0.005  {\rm nm}^{-2}$.
  The location of the lateral bond under maximal force within the GTP-cap 
  is shown in figures \ref{fig:HP_vec_05} and \ref{fig:HP_vec_0005}
  for strong and weak bonds, respectively, as thick black line. 
}
 \label{fig:kraft_vec}
\end{figure}

Here we want to take a different approach and ask, which rupture 
forces can initiate  catastrophe-like  events within our 
mechanical model of the MT.
To answer this question we consider the most probable hydrolysis 
pathway and calculate the maximal lateral bond force $F_{\rm max}$ 
within  the MT lattice at every hydrolysis step.
The results are shown in figure \ref{fig:kraft_random} for 
random hydrolysis and figure  \ref{fig:kraft_vec} for 
vectorial hydrolysis. 
The location of the lateral bond under maximal force within the GTP-cap 
  is shown in the corresponding 
  figures \ref{fig:HP_random_05}--\ref{fig:HP_vec_0005}
  as thick black line.

For random hydrolysis, mechanical forces favor hydrolysis 
at the last layer of the  GTP-cap towards the plus end.
For random hydrolysis and strong lateral bonds, see figure
\ref{fig:kraft_random}a,  we see a strong increase in the maximal 
bond force  
$F_{\rm max}$  after hydrolysis of $n_{\rm hyd} = 14$ dimers. 
At this point, the first dimer in the second to last layer
is hydrolyzed after complete hydrolysis of the last layer 
at the plus end. For weak lateral bonds,  see figure
\ref{fig:kraft_random}b, we even see a periodic layer-by-layer pattern
in the increase of $F_{\rm max}$ with pronounced 
jumps at $n_{\rm hyd} = 1, 14, 27$, where hydrolysis reaches the 
next layer starting from the plus end. 
In the corresponding figures \ref{fig:HP_random_05} for strong lateral 
bonds and \ref{fig:HP_random_0005} for weak lateral bonds, 
the location of the maximally strained
lateral bond is shown as thick black line.
The maximally strained bond is at the plus end layer of the 
GTP-cap bond next to  the same protofilament as the hydrolyzed dimer. 
Hydrolysis of the first dimer in a layer increases the 
spontaneous curvature of that protofilament and, thus, 
increases the 
strain on lateral bonds of the same protofilament in the last layer.

For vectorial hydrolysis, only dimers at the GTP-GDP
  interface can be hydrolyzed, and hydrolysis has to advance from the 
minus end side. 
The results for vectorial  hydrolysis and  strong 
lateral bonds, see  figure 
 \ref{fig:kraft_vec}a,  show that after hydrolysis of $n_{\rm hyd} = 16$ dimers,
when the third protofilament becomes completely hydrolyzed,
there is a pronounced step-like  increase in $F_{\rm max}$. 
Similarly, for weak lateral bonds, see figure 
 \ref{fig:kraft_vec}b, there is a strong increase 
 after hydrolysis of $n_{\rm hyd} = 5$ dimers,  when the first 
protofilament becomes completely hydrolyzed.  
A completely hydrolyzed protofilament prefers to 
assume its  curved 
equilibrium state and, thus,  exerts strong outward forces on neighboring 
 stabilizing protofilaments with GTP-caps. 
Therefore, with each completely hydrolyzed protofilament 
the  strain on  the lateral bonds increases. 
For weak bonds a single hydrolyzed protofilament is sufficient 
to cause a pronounced increase in $F_{\rm max}$, whereas for 
strong bonds several hydrolyzed protofilaments seem to be necessary. 
In the corresponding figures \ref{fig:HP_vec_05} for strong lateral 
bonds and \ref{fig:HP_vec_0005} for weak lateral bonds 
the location of the maximally strained
lateral bond is shown as thick black line: the bond under maximal 
force occurs at the last layer of the cap next to the 
completely hydrolyzed protofilaments. 

Typical values for  $F_{\rm max}$ depend on the strength of lateral 
bonds. For strong lateral bonds ($k/\kappa=0.5  {\rm nm}^{-2}$) we find 
$F_{\rm max}/\kappa \sim 0.06 {\rm nm}^{-1}$ both for random and vectorial 
hydrolysis.  For $\kappa \sim 25 k_BT$ well within the 
considered range $14 k_BT \le \kappa \ll 68 k_BT$ of dimer bending rigidities,
this gives maximal forces $F_{\rm max} \sim 6{\rm pN}$. 
For weak lateral bonds ($k/\kappa=0.005  {\rm nm}^{-2}$) we find 
smaller maximal forces 
$F_{\rm max}/\kappa \sim 0.01 {\rm nm}^{-1}$ corresponding to 
$F_{\rm max} \sim 1{\rm pN}$, which are again  similar 
 for random and vectorial hydrolysis.

%%%%%%%%%%%%%%%%%%%%%%
\subsubsection{Bond rupture and catastrophe initiation.}

To further characterize the susceptibility to a mechanical instability,
i.e., a rupture avalanche 
as it happens in catastrophes, we also investigated the 
{\it change} $\Delta F_{\rm max}$ in the
maximal lateral bond force  if we rupture the 
maximally strained bond and calculate the new maximal 
lateral bond force on an intact bond, 
see the lower plots in figure \ref{fig:kraft_random} for 
random hydrolysis and figure  \ref{fig:kraft_vec} for 
vectorial hydrolysis. 
If $\Delta F_{\rm max}$ is large and positive, there is an 
 increase in the maximal bond force and the possibility 
of an instability: a rupture force 
$F_{\rm rup}<F_{\rm max}$ 
below this maximal force level  can  give rise to 
continued bond rupture. 
Vice versa, a negative value for  $\Delta F_{\rm max}$  signals a 
mechanical stable situation. 
Remarkably, we find a strong correlation between the pronounced 
 increase
in $F_{\rm max}$ and a pronounced  increase in $\Delta F_{\rm max}$ to 
larger positive values,
see figure \ref{fig:kraft_random} for 
random hydrolysis and figure   \ref{fig:kraft_vec} for 
vectorial hydrolysis.

This suggests that the pronounced 
 increase in the maximal lateral bond force 
is the starting point of a catastrophe-like rupture avalanche,
if we choose the value for the rupture force $F_{\rm rup}$ 
such that after the  increase in $F_{\rm max}$, the maximal 
lateral bond force exceeds the rupture threshold.
Using this criterion, a reasonable value for the rupture 
force   for random hydrolysis is 
$F_{\rm rup}\sim 0.062 \kappa {\rm nm}^{-1}$ 
for strong lateral bonds and $F_{\rm rup}\sim 0.011 \kappa {\rm nm}^{-1}$
for weak lateral bonds.
For vectorial hydrolysis, 
this suggests rupture forces $F_{\rm rup}\sim 0.055 \kappa {\rm nm}^{-1}$ 
for strong lateral bonds and $F_{\rm rup}\sim 0.0125 \kappa {\rm nm}^{-1}$
for weak lateral bonds. 
With values of $\kappa\sim 25 k_BT$ 
from the range 
$14 k_BT \le \kappa \ll 68 k_BT$ according to the bounds 
(\ref{eq:kappa_upperbound}) and (\ref{eq:kappa_lowerbound}),
the resulting rupture forces are $F_{\rm rup}\sim 
5-6{\rm pN}$ for strong bonds or $1.2-1.3 {\rm pN}$ for weak bonds, 
respectively. 

\begin{figure}
 \includegraphics[width=0.8\textwidth]{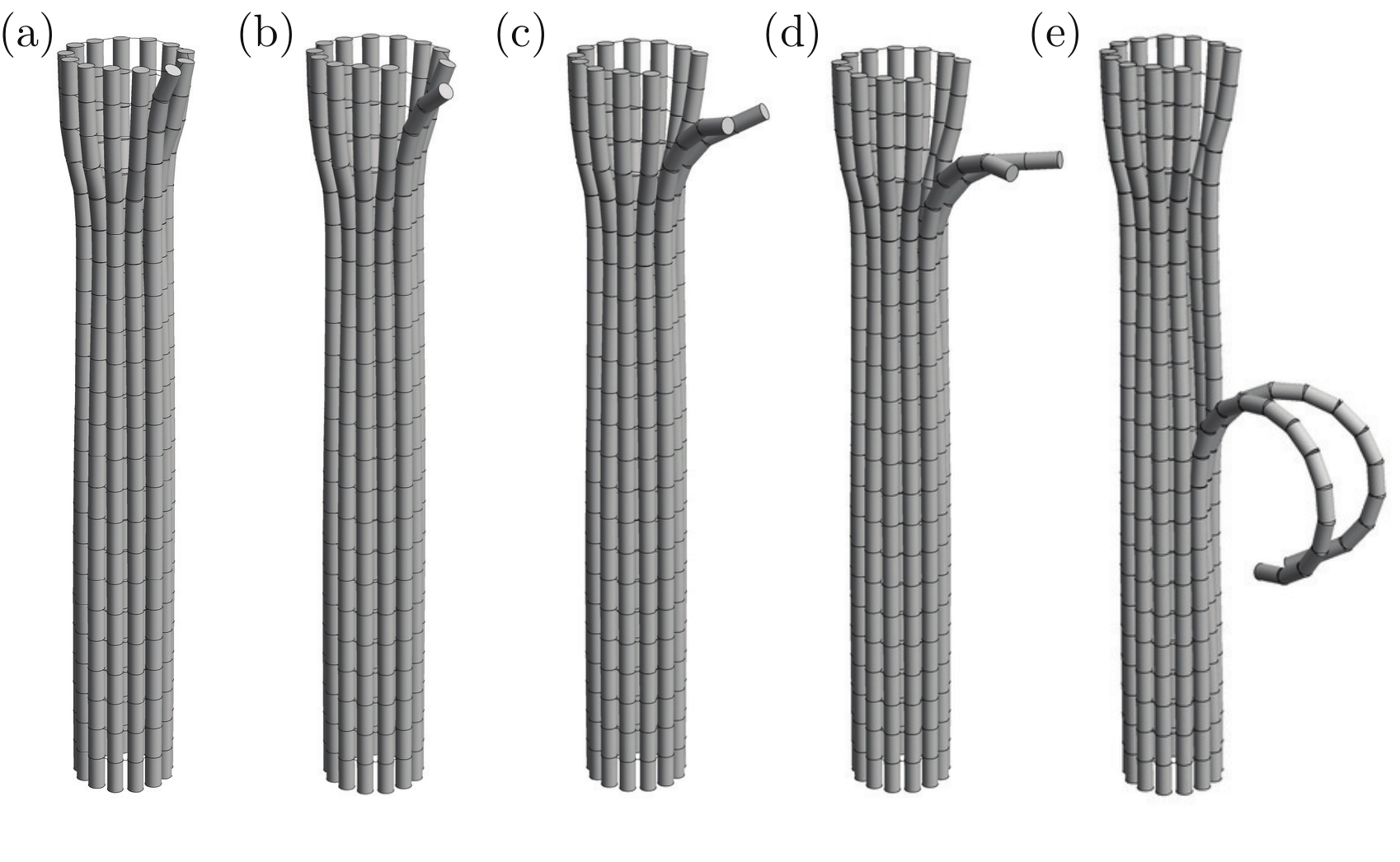}
\caption{
   Catastrophe event for vectorial  hydrolysis and weak 
   lateral bond springs $k/\kappa=0.005  {\rm nm}^{-2}$ using 
  a rupture force $F_{\rm rup}= 0.0125 \kappa {\rm nm}^{-1}$
  ($m_1=20$ and $m_2=3$).
  The numbers $n_{\rm hyd}$ of hydrolyzed dimers and 
$n_{\rm cut}$ of cut lateral bonds are 
(a) $n_{\rm hyd}=5$, $n_{\rm cut}=1$, (b) $n_{\rm hyd}=6$, $n_{\rm cut}=4$, 
(c) $n_{\rm hyd}=6$, $n_{\rm cut}=8$, (d) $n_{\rm hyd}=6$, $n_{\rm cut}=11$,
   (e) $n_{\rm hyd}=6$, $n_{\rm cut}=33$.
}
 \label{fig:catastrophe_vec}
\end{figure}

\begin{figure}
 \includegraphics[width=0.8\textwidth]{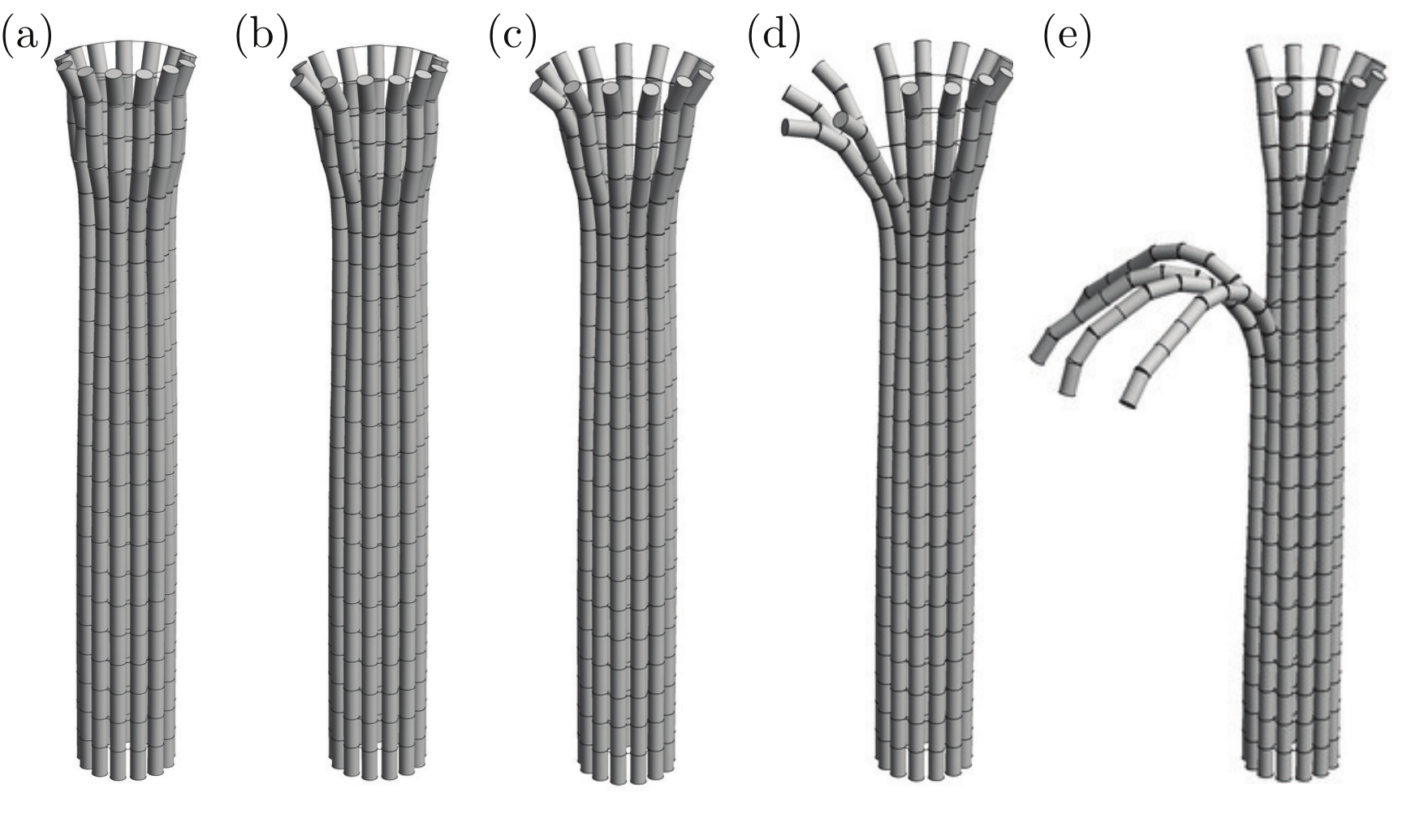}
\caption{
   Catastrophe event for random  hydrolysis and weak 
   lateral bond springs $k/\kappa=0.005  {\rm nm}^{-2}$ using 
  a rupture force $F_{\rm rup}= 0.011 \kappa {\rm nm}^{-1}$
  ($m_1=20$ and $m_2=3$).
  The numbers $n_{\rm hyd}$ of hydrolyzed dimers and 
$n_{\rm cut}$ of cut lateral bonds are 
(a) $n_{\rm hyd}=14$, $n_{\rm cut}=1$, (b) $n_{\rm hyd}=14$, $n_{\rm cut}=6$, 
(c) $n_{\rm hyd}=14$, $n_{\rm cut}=15$, (d) $n_{\rm hyd}=14$, $n_{\rm cut}=24$,
   (e) $n_{\rm hyd}=14$, $n_{\rm cut}=45$.
}
 \label{fig:catastrophe_random}
\end{figure}

We test whether our model exhibits catastrophe-like
events if we fix a value for the rupture force according to this 
criterion by  including bond rupture into our 
algorithm for the most probable hydrolysis pathway.
After each hydrolysis step we check
 whether any lateral bond spring is loaded by a force larger 
than the rupture force. If this is the case, 
 we rupture the corresponding spring by cutting it 
and mechanically relax the MT lattice. Then we test 
again for bond rupture and so on. If no further rupture 
events occur, we continue with the next hydrolysis step.

Using this simulation procedure we indeed find catastrophe events with
continued bond rupture similar to experimentally observed catastrophes.  For
vectorial hydrolysis, see figure \ref{fig:catastrophe_vec}, typically single
protofilaments peel off the MT lattice and curl into ``ram's horn''
conformations.  During all the rupture events shown in figure
\ref{fig:catastrophe_vec}, we have $n_{\rm hyd}=5$ or $6$, i.e., hydrolysis
practically stops during such an event.  
The hydrolysis state at $n_{\rm  hyd}=5$ 
consists of a single completely hydrolyzed protofilament (see figure
\ref{fig:HP_vec_0005}). Therefore, the initial stress distribution in lateral
bonds is concentrated on this protofilament.
The stress due to the preferred curved configuration 
of this  protofilament triggers the bond rupture of neighboring 
lateral bonds, and this protofilament starts to peel off 
during the  catastrophe.

For random hydrolysis, see figure \ref{fig:catastrophe_random},
 typically several protofilaments
peel off the MT lattice. Also for this hydrolysis mechanism, the hydrolysis 
stops during such an event: the entire 
catastrophe event in figure \ref{fig:catastrophe_random} happens 
at $n_{\rm hyd}=14$.
For random hydrolysis the 
 stress is typically distributed among the entire top layer of 
lateral bonds because 
mechanically dominated hydrolysis proceeds layer by layer;
the state  $n_{\rm hyd}=14$  consists 
of a completely hydrolyzed plus end layer of dimers with one 
additional hydrolysis event in the next layer, see 
figure \ref{fig:HP_random_0005}. This gives rise to several 
bond rupture events within the top layer and several protofilaments
peeling off the MT. 

These different catastrophe characteristics for vectorial and 
random hydrolysis  could be an interesting issue 
for future experimental studies.

%%%%%%%%%
\section{Discussion and Conclusion}

We introduced a 
mechanical model, which gives stable tubular MT structures
if a stabilizing GTP-cap is present
in accordance with experimental observations. 
The model includes intra-dimer bending and  one lateral bond per dimer. 
In order to avoid  overlapping of tubulin dimers we also include a 
sufficiently strong hard core interaction. 
We use the allosteric model for the hydrolysis state of 
tubulin dimers: GTP dimers are straight; after hydrolysis, a GDP dimer 
has an equilibrium bending angle of $22^\circ$. 

We obtained several constraints 
for the model parameters, the bending rigidity $\kappa$ 
and the lateral bond strength $k$: 
(i) The ratio $k/\kappa$ is constrained 
by a lower bound $k/\kappa \ge 0.001 {\rm nm}^{-2}$, see 
 (\ref{eq:klowerbound}), which ensures that 
GTP-capped MTs do not spontaneously 
acquire a strongly bent shape similar to the 
 ram's horn configuration.
(ii) The value for $\kappa$ is constrained by an upper bound 
(\ref{eq:kappa_upperbound}) because hydrolysis is not observed to 
be a reversible reaction and the free energy released in hydrolysis should 
 exceed the mechanical energy increase in the MT lattice during hydrolysis.
(iii) If we additionally 
assume that hydrolysis is dominated by mechanical forces 
in the MT lattice, i.e., typical mechanical energy changes of the  the MT
lattice during hydrolysis exceed the thermal energy $k_BT$, the 
value for $\kappa$ is also constrained by a lower bound 
(\ref{eq:kappa_lowerbound}).
The bounds (ii) and (iii) define a range 
$14 k_BT \le \kappa \ll 68 k_BT$ of dimer bending rigidities.

The mechanical model 
 allows us to investigate the interplay of mechanical forces in the 
MT lattice and hydrolysis, which has not been done previously. 
The interaction via the mechanics of the MT lattice 
can give rise to possible correlation effects in the hydrolysis 
dynamics, which have not been taken into account before. 
Under the assumption of a mechanically dominated hydrolysis reaction
the concept  of a {\it most probable hydrolysis pathway} becomes 
very useful. We calculated  most probable hydrolysis pathways numerically
both for random hydrolysis 
(figures \ref{fig:HP_random_05} and \ref{fig:HP_random_0005}),
where all GTP-dimers in the MT cap can be hydrolyzed,
and for vectorial hydrolysis (figures \ref{fig:HP_vec_05} and
\ref{fig:HP_vec_0005}), where only GTP-dimers at the GTP-GDP interface
can be hydrolyzed. 
We also studied the effect of lateral bond strength on the hydrolysis 
pathway.

For mechanically dominated 
random hydrolysis, the most probable hydrolysis pathway 
shows a preference for layer-by-layer hydrolysis starting 
at the GTP-layer at the plus end. 
For mechanically dominated 
vectorial  hydrolysis, the most probable hydrolysis pathway 
shows a preference for hydrolysing complete protofilaments 
towards the plus end starting from 
 a certain ``nucleus'' configuration consisting of three
neighboring hydrolyzed GDP-dimers within the same layer.
This could be related to the experimental observation of 
several rate-limiting steps for catastrophe 
initiation \cite{Gardner2011a}.

We also investigated the lateral bond forces occurring during hydrolysis.
For random hydrolysis, mechanically dominated hydrolysis proceeds in a 
layer-by-layer fashion starting at the plus end, 
and we found sharp increases of the maximal 
lateral bond force if hydrolysis of one layer has been completed 
and hydrolysis of the next layer starts. 
The respective maximal forces occur  at the 
plus end of the MT.
For vectorial hydrolysis and weak lateral bonds, we 
 found a sharp increase of the maximal lateral bond force if a 
single protofilament becomes completely hydrolyzed.
For stronger lateral bonds, several protofilaments need to become 
completely hydrolyzed to trigger a similar increase in the maximal 
lateral bond force. The respective maximal forces occur right at the 
plus end. 

Moreover, we observe that  the MT lattice becomes mechanically unstable 
at the sharp increase: rupture of  the maximally strained bond
further {\it increases} the maximal bond force, which signals an 
instability with respect to a catastrophe event initiated by 
lateral bond rupture. 
If the  rupture force value 
lies within the force range set by the sharp increase of the 
maximal lateral bond force, we indeed find continued bond rupture 
and catastrophe events as observed in experiments, see 
figures \ref{fig:catastrophe_vec} and \ref{fig:catastrophe_random}.  
Using this criterion we find  rupture force values 
between 1pN for weak and 5pN for strong lateral bonds.
For vectorial hydrolysis, we find catastrophes starting 
 with single protofilaments
peeling off the MT. For random hydrolysis, on the other hand,  
we typically see several protofilaments peeling off the MT. 
These characteristic differences in our simulations 
could motivate further 
experimental studies of this issue. 

Our results suggest several routes for future work.
Firstly, we only studied MTs of fixed length for simplicity. 
Further investigations will include stochastic  polymerization
and depolymerization similar to the models  in 
\cite{VanBuren2002,VanBuren2005}.
Secondly, we assumed so far that hydrolysis is mechanically 
dominated and forces on the most probable hydrolysis pathway by 
selecting the next GTP-dimer to be hydrolyzed according to the 
maximal mechanical energy gain. 
Future models should be fully stochastic with 
hydrolysis rates {\it modulated}
by mechanical energies in order to include all hydrolysis pathways 
with their respective statistical weight into the analysis.
Finally, a similar mechanical MT model and its coupling to 
hydrolysis should be investigated not only for the allosteric model
but also for the lattice model of dimer hydrolysis.

%%%%%%%%%%%%%%%%
\ack

We thank Bj\"orn Zelinski and Sebastian Knoche for 
fruitful discussions,  and  we
acknowledge support by  the Deutsche Forschungsgemeinschaft 
(KI 662/4-1).

%%%%%%%%%%%%%%%%%%%%%%%%%%%%%%


\section*{References}
\begin{thebibliography}{10}

\bibitem{Mitchison2001}
Mitchison T J and Salmon E D 2001, 
%Mitosis: a history of division
{\it Nat. Cell Biol.} {\bf 3}, E17--E21


\bibitem{D05}
Dogterom M, Kerssemakers J W J, Romet-Lemonne G and  Janson M E 2005, 
%Force generation by dynamic microtubules.
{\it Curr. Opin. Cell Biol.} {\bf 17}, 67--74


\bibitem{daga2006}
Daga R R, Yonetani A and Chang F 2006, 
%Asymmetric microtubule pushing forces in nuclear centering.
{\it Curr. Biol.} {\bf 16}, 1544--50


\bibitem{SD07}
Siegrist S E and Doe C Q 2007, 
%Microtubule-induced cortical cell polarity.
{\it Genes Dev.} {\bf 21}, 483--96


\bibitem{Picone2010}
Picone R, Ren X, Ivanovitch K D, Clarke J D W, McKendry R A, and 
Baum B 2010, 
%A polarised population of dynamic microtubules mediates homeostatic length control in animal cells.
{\it PLoS Biol.} {\bf 8}, e1000542 


\bibitem{Dehmelt2003}
Dehmelt L, Smart F M, Ozer R S and Halpain S 2003, 
%The role of microtubule-associated protein 2c in the reorganization of microtubules and lamellipodia during neurite initiation.
{\it J. Neurosci.} {\bf 23}, 9479--90


\bibitem{Mitch1984}
Mitchison T and Kirschner M 1984, 
%Dynamic instability of microtubule growth
{\it Nature} {\bf 312}, 237--42


\bibitem{Kirschner1974}
Kirschner M, Williams R, Weingarten M and Gerhart J C 1974,
%Microtubules from mammalian brain: some properties of their depolymerization products and a proposed mechanism of assembly and disassembly.
{\it Proc. Natl. Acad. Sci. USA} {\bf 71}, 1159--63


\bibitem{Wang2005}
Wang H and Nogales E 2005,
%Nucleotide-dependent bending flexibility of tubulin regulates microtubule assembly.
{\it Nature} {\bf 435}, 911--5


\bibitem{Buey2006}
Buey R, Diaz J, and Adreu J 2006,
%The nucleotide switch of tubulin and microtubule assembly: a
%polymerization-driven structural change.
{\it Biochemistry} {\bf 45}, 5933--8


\bibitem{Wu2012}
Wu Z, Nogales E and Xing J 2012,
%Comparative studies of microtubule mechanics with two competing models
%are separatedsuggest functional roles of alternative tubulin lateral
%are separatedinteractions.
{\it Biophys. J.} {\bf 102}, 2687--96


\bibitem{nogales2006}
Nogales E and Wang H 2006,
%Structural mechanisms underlying nucleotide-dependent self-assembly of tubulin and its relatives.
{\it Curr. Opin. Struct. Biol.} {\bf 16}, 221--9 


\bibitem{Mueller1998}
M\"uller-Reichert T, Chr{\'e}tien D, Severin F and
Hyman A A 1998,
%Structural changes at microtubule ends accompanying GTP hydrolysis: 
%information from a slowly hydrolyzable analogue of GTP, guanylyl
% (alpha,beta)methylenediphosphonate.
{\it Proc. Natl. Acad. Sci. USA} {\bf 95}, 3661--6


\bibitem{Gardner2011a}
Gardner M K,  Zanic M,  Gell C,  Bormuth V and Howard J 2011,
%Depolymerizing Kinesins Kip3 and MCAK Shape Cellular Microtubule Architecture by Differential Control of Catastrophe
{\it Cell} {\bf 147}, 1092--103


\bibitem{Doorn2000}
Van Doorn G, Tanase C, Mulder B M and Dogterom M 2000,
%On the stall force for growing microtubules.
{\it Eur. Biophys. J.} {\bf 29}, 2--6


\bibitem{Kolomeisky2001}
Kolomeisky A B and Fisher M 2001,
%Force-velocity relation for growing microtubules.
{\it Biophys. J.} {\bf 80}, 149--54


\bibitem{Stukalin2004} 
Stukalin E B and Kolomeisky A B 2004,
%Simple growth models of rigid multifilament biopolymers.
{\it J. Chem. Phys.} {\bf 121}, 1097--104

\bibitem{Ranjith2009}
Ranjith  P, Lacoste  D, Mallick K and Joanny J-F 2009,
%Nonequilibrium self-assembly of a filament coupled to ATP/GTP hydrolysis.
{\it Biophys. J.} {\bf 96}, 2146--59

\bibitem{Krawczyk2011}
Krawczyk J and Kierfeld J 2011,
%Stall force of polymerizing microtubules and filament bundles
{\it EPL} {\bf 93}, 28006


\bibitem{Dogterom1993}
Dogterom M and Leibler S 1993,
%Physical aspects of the growth and regulation of microtubule structures
{\it Phys. Rev. Lett.} {\bf 70}, 1347--50


\bibitem{Flyvbjerg1994}
Flyvbjerg H, Holy T and Leibler S 1994,
%Stochastic Dynamics of Microtubules: A Model for Caps and Catastrophes
{\it Phys. Rev. Lett.} {\bf 73}, 2372--5


\bibitem{Flyvbjerg1996}
Flyvbjerg H, Holy T and Leibler S 1996,
%Microtubule dynamics: Caps, catastrophes, and coupled hydrolysis.
{\it Phys. Rev. E} {\bf 54}, 5538--60


\bibitem{Zelinski2012}
Zelinski B, M\"uller N  and Kierfeld J 2012, 
%Dynamics and length distribution of microtubules under force and confinement
{\it Phys. Rev. E} {\bf 86}, 041918


\bibitem{Zelinski2013}
Zelinski B and Kierfeld J 2013, 
%Cooperative dynamics of microtubule ensembles: 
%Polymerization forces and rescue-induced oscillations
{\it Phys. Rev. E} {\bf 87}, 012703


\bibitem{Sept2010}
Sept D and MacKintosh F C 2010,
%Microtubule Elasticity: Connecting All-Atom Simulations with Continuum Mechanics
{\it Phys. Rev. Lett.} {\bf 104}, 018101

\bibitem{Wells2010}
%Mechanical properties of a complete microtubule 
%revealed through molecular dynamics simulation.
Wells D B and Aksimentiev A 2010,
{\it Biophys. J.} {\bf 99}, 629--37

\bibitem{Grafmuller2011}
Grafm\"{u}ller A and Voth G 2011,
%Intrinsic bending of microtubule protofilaments.
{\it Structure} {\bf 19}, 409--17



\bibitem{Grafmuller2013}
Grafm\"{u}ller A, Noya E G and Voth G 2013,
%Nucleotide-dependent lateral and longitudinal interactions in microtubules.
{\it J. Mol. Biol.} {\bf 425}, 2232--46


\bibitem{Bayley1990}
Bayley  P M, Schilstra M J and Martin S R 1990,
%Microtubule dynamic instability: numerical simulation of microtubule transition properties using a Lateral Cap model.
{\it J. Cell Sci.} {\bf 95} 33--48

\bibitem{VanBuren2002}
VanBuren V,  Odde D J and Cassimeris L 2002,
%Estimates of lateral and longitudinal bond energies within the microtubule are separated lattice.
{\it Proc. Natl. Acad. Sci. USA} {\bf 99}, 6035--40


\bibitem{Brun2009a}
Brun L,  Rupp B, Ward J J and N\'{e}d\'{e}lec F 2009,
%A theory of microtubule catastrophes and their regulation.
{\it Proc. Natl. Acad. Sci. USA} {\bf 106}, 21173--8


\bibitem{Bowne2013}
Bowne-Anderson H, Zanic M, Kauer M and Howard J 2013
%Microtubule dynamic instability: a new model with coupled GTP hydrolysis and
%multistep catastrophe.
{\bf Bioessays} {\bf 35}, 452--61

\bibitem{Jemseena2013}
Jemseena V and Gopalakrishnan M 2013,
%Microtubule catastrophe from protofilament dynamics
{\it Phys. Rev. E} {\bf 88}, 032717


\bibitem{Padinhateeri2012}
%Random hydrolysis controls the dynamic instability of microtubules.
Padinhateeri R, Kolomeisky, A B and Lacoste, D 2012,
{\it Biophys. J.} {\bf 102}, 1274--83


\bibitem{Li2013}
Li X L and Kolmeisky A B 2013, 
%Theoretical analysis of microtubules dynamics using a 
%physical-chemical description of hydrolysis.
{\it J. Phys. Chem. B} {\bf 117}, 9217--23



\bibitem{Li2009}
Li X,  Kierfeld J and Lipowsky R 2009,
%Actin Polymerization and Depolymerization Coupled to Cooperative Hydrolysis
{\it Phys. Rev. Lett.} {\bf 103}, 048102


\bibitem{Li2010}
Li X,   Lipowsky R and Kierfeld J 2010,
%Coupling of actin hydrolysis and polymerization: Reduced description with two nucleotide states
{\it  EPL} {\bf 89}, 38010 


\bibitem{Drechsel1994}
Drechsel D and Kirshner M 1994,
%The minimum GTP cap required to stabilize microtubules.
{\it Curr. Biol.} {\bf 4}, 1053--61


\bibitem{Desai1997}
Desai A and Mitchison T J 1997,
%Microtubule polymerization dynamics
{\it Annu. Rev. Cell Dev. Biol.} {\bf 13}, 83--117


\bibitem{Schek2007}
Schek H T, Gardner M K, Cheng J,  Odde D J and Hunt A J 2007,
%Microtubule assembly dynamics at the nanoscale.
{\it Curr. Biol.} {\bf 17}, 1445--55


\bibitem{Molodtsov2005}
Molodtsov M I,  Ermakova A E, Shnol E E, 
Grishchuk E L, McIntosh, J R and
Ataullakhanov F I 2005, 
%A molecular-mechanical model of the microtubule.
{\it Biophys. J.} {\bf 88}, 3167--79


\bibitem{Mohrbach2010}
Mohrbach H, Johner A, and Kuli\'{c} I M 2010,
%Tubulin Bistability and Polymorphic Dynamics of Microtubules
{\it Phys. Rev. Lett.} {\bf 105}, 268102


\bibitem{VanBuren2005}
VanBuren V,  Cassimeris L and Odde D J 2005,
%Mechanochemical model of microtubule structure and self-assembly kinetics.
{\it Biophys. J.} {\bf 89}, 2911--26


\bibitem{Carlier1987}
Carlier M F, Didry D and Pantaloni S 1987,
%Microtubule elongation and guanosine 5'-triphosphate hydrolysis. Role of guanine nucleotides in microtubule dynamics.
{\it Biochemistry} {\bf 26}, 4428--37


\bibitem{Nogales2000}
Nogales E 2000,
%Structural insights into microtubule function.
{\it Annu. Rev. Biochem.} {\bf 69}, 277--302


\bibitem{Kis2008}
Kis A, Kasas S, Kulik A J, Catsicas S and Forro L 2008,
%Temperature-dependent elasticity of microtubules.
{\it Langmuir} {\bf 24}, 6176--81


\bibitem{Molodtsov2005b}
Molodtsov M I, Grishchuk E L,   McIntosh, J R and
Ataullakhanov F I 2005, 
%Force production by depolymerizing microtubules: a theoretical study.
{\it Proc. Natl. Acad. Sci. USA} {\bf 102}, 4353--8

\bibitem{Pampaloni}
Pampaloni F, Lattanzi G, Jonas A, Surrey T,  Frey E and 
Florin, E L 2006, 
%Thermal fluctuations of grafted microtubules provide 
%evidence of a length-dependent persistence length.
{\it Proc. Natl. Acad. Sci. USA} {\bf 103}, 10248--53

\bibitem{Bell}
Bell G I 1978,
{\it Science} {\bf 200}, 618--627



\end{thebibliography}
\end{document}